\documentclass[twocolumn,noabbrev]{aa}
\usepackage{amsmath}
\usepackage{mathrsfs}
\usepackage{comment}
\usepackage{natbib,twoopt}
\usepackage[hyphenbreaks]{breakurl}
\usepackage[breaklinks]{hyperref}      %% to avoid \citeads line fills, add "draft"

\usepackage{cleveref}
\usepackage{subfigure}
\usepackage[dvipsnames]{xcolor}
\usepackage{placeins}
\usepackage{capt-of}
\usepackage{linenoaa}
\usepackage{graphicx}
\usepackage[percent]{overpic}
\usepackage{tikz}
\usepackage{savesym}
\savesymbol{tablenum}
\usepackage{siunitx}
\restoresymbol{SIX}{tablenum}
\DeclareSIUnit{\erg}{erg}

\newcommand{\dmnote}[1]{{\bf \color{blue}[D.M.: #1]}}

\newcommand{\St}{{\rm St}}

\crefalias{pequation}{equation}
\crefformat{pequation}{(eq.~#2#1#3)}
\Crefformat{pequation}{(Equation~#2#1#3)}

\bibpunct{(}{)}{;}{a}{}{,}             %% natbib format for A&A and ApJ
\makeatletter
  \newcommandtwoopt{\citeads}[3][][]{\href{http://adsabs.harvard.edu/abs/#3}%
    {\def\hyper@linkstart##1##2{}%
     \let\hyper@linkend\@empty\citealp[#1][#2]{#3}}}
  \newcommandtwoopt{\citepads}[3][][]{\href{http://adsabs.harvard.edu/abs/#3}%
    {\def\hyper@linkstart##1##2{}%
     \let\hyper@linkend\@empty\citep[#1][#2]{#3}}}
  \newcommandtwoopt{\citetads}[3][][]{\href{http://adsabs.harvard.edu/abs/#3}%
    {\def\hyper@linkstart##1##2{}%
     \let\hyper@linkend\@empty\citet[#1][#2]{#3}}}
  \newcommandtwoopt{\citeyearads}[3][][]%
    {\href{http://adsabs.harvard.edu/abs/#3}
    {\def\hyper@linkstart##1##2{}%
     \let\hyper@linkend\@empty\citeyear[#1][#2]{#3}}}
\makeatother

\begin{document}

\title{Simulating the thermodynamics of gas, radiation, and multispecies dust}
\subtitle{Method and applications to protoplanetary disks}

   \author{Dhruv Muley \inst{1, 2}
        \and David Melon Fuksman \inst{1}
        \and Prakruti Sudarshan \inst{1}
        \and Alexandros Ziampras \inst{3}
        \and Mario Flock \inst{1}
        }

   \institute{Max-Planck-Institut f\"ur Astronomie, Königstuhl 17, Heidelberg, DE 69117
    \and
    Max-Planck-Institut f\"ur Astrophysik, Karl-Schwarzschild-Straße 1, Garching bei M\"unchen, DE 85748\\
    \email{\href{mailto:dmuley@mpa-garching.mpg.de}{dmuley@mpa-garching.mpg.de}}
    \and
    Universit\"atssternwarte M\"unchen, LMU\\
              }
   \date{Received 26 June 2026; accepted 21 September 2026}
\abstract{The thermodynamics of protoplanetary disks, governed by a complex interplay between gas, dust, and radiation, exerts a strong influence on their morphology and dynamics at both large and small scales. Historically, hydrodynamical simulations have treated thermodynamics with approximate prescriptions, such as local isothermality, parametrized $\beta$-cooling, and radiation hydrodynamics with gas-dust thermal equilibrium. However, a more comprehensive and self-consistent approach would more accurately reproduce the wealth of features found in observations, which in modern times have reached unprecedented spectral, angular, and vertical resolution. With this motivation, we have devised a radiation-hydrodynamics scheme for the PLUTO code with energy exchange (via absorption, emission, and collisions) for gas, radiation, and multiple species of dust. Dust-gas dynamics are handled by treating each dust species as a pressureless, diffusive fluid, ensuring that our scheme can represent changes in disk illumination caused by grain settling and trapping. We demonstrate the effectiveness of our scheme in several test problems, and conclude by discussing its relevance to open questions in protoplanetary disk studies.}


   \maketitle
   \nolinenumbers

\section{Introduction} \label{sec:intro}
Circumstellar disks around nascent stars consist of both gas and dust, which interact with one another primarily via collisions between gas molecules and dust grains. Although the dust component constitutes only a small fraction of the disk mass (typically taken to be ${\sim}$1\%, following interstellar-medium values), it nevertheless exert a profound influence on the dynamical and thermodynamical evolution of the disk. Larger, millimeter-sized grains, for which collisional coupling times $t_s$ are longer, experience comparatively weak drag forces from the gas. This causes them to settle to the midplane \citep[e.g.,][]{Morfill1984,Fromang2009,Dullemond2022} and concentrate at radial pressure maxima \citep[][]{Pinilla2012,Dullemond2018}, giving rise to bright, observable rings in the millimeter continuum \citep[e.g.,][and other papers in series]{Andrews2018} and encouraging planetesimal formation via the streaming instability \citep{Youdin2005}. Smaller grains, with lower $t_s$, experience stronger drag and remain entrained in the gas. In the upper layers of the disk, they absorb stellar irradiation and reradiate it toward the midplane. Given that the frequency-integrated gas opacity is typically low compared to that of the dust, the dust temperature often functions as a background profile to which the gas temperature relaxes via collision-mediated thermal accommodation \citep{BurkeHollenbach83}. In the disk midplane, gas-grain collisions are frequent and the dust and gas temperatures rapidly come to thermal equilibrium; in the atmosphere, they are comparatively rare, and  dust and gas temperatures can decouple.

To model disk thermodynamics, the most comprehensive approaches are the thermochemical models, such as the Protoplanetary Disk Model \citep[ProDiMo; ][]{Woitke2009} and Disks And Lines \citep[DALI; ][]{Bruderer2012}. Such schemes typically include gas, a multi-size dust distribution, a multi-wavelength radiation field, and a chemical network (with e.g, $^{12}$CO, $^{13}$CO, H$_2$S), capturing the exchange of mass and energy between these species as they absorb, emit, collide, and chemically react with one another. Despite their utility and realism, the equilibrium assumptions and substantial computational cost of such methods make them impractical to couple with gas dynamics, where energy exchange must be recomputed at each computational timestep. For this reason, simulations have historically assumed a locally isothermal \citep[e.g.,][]{Dangelo2010,Dong2016,Juhasz2018} equation of state, or an adiabatic equation of state with exponential thermal relaxation to a fixed background, over some timescale $t_c \equiv \beta\Omega^{-1}$ \citep[$\beta$-cooling; see for example][]{Gammie2001,Zhu2015,Muley2021,Zhang2024}.

In recent years, a number of analytical and numerical studies have demonstrated that the outcomes of disk-planet interaction \citep[e.g.,][]{Miranda2019,Miranda2020,Ziampras2023} and disk fluid instabilities \citep[e.g.,][]{Klahr2014,Manger2021,Pfeil2024} depend sensitively on temperature structure and thermal relaxation timescales. This has encouraged the use of physically-motivated thermodynamic prescriptions in hydrodynamical simulations \citep[e.g.,][]{Bae21,Ziampras2026}, based on detailed calculations of equilibrium dust distribution, temperature structure, and collisional/radiative relaxation timescales. With live radiative transfer, computed using techniques such as flux-limited diffusion \citep[FLD; ][]{Levermore1981,Kley1989}, M1 \citep{Levermore1984}, and discrete ordinates \citep[e.g.,][]{Jiang2021}, the background temperature can be made to evolve dynamically, a useful property for the study of shadowed disks \citep[e.g.,][]{Montesinos2016,Zhang2024_1,Muley2024b} or luminous, accreting protoplanets \citep{Muley2024}, among other applications. The three-temperature (3T) scheme of \cite{Muley2023}, based on the M1 method developed by \citep{MelonFuksman2019,MelonFuksman2021} for the PLUTO hydrodynamics code \citep{Mignone2007}, self-consistently evolves the separate energies of gas, dust, and radiation via thermal accommodation, absorption/emission of radiation, and stellar irradiation. In line with physical expectations, this enables the dust and gas temperatures to decouple in the upper layers of the disk.

None of these methods considers the independent dynamics of the dust. For this problem, the literature contains two major approaches. Modeling dust as superparticles
\citep[e.g.,][]{Klahr1997,Johansen2005,Bai2010,Hopkins2016,Yang2016} allows the treatment of all dust grain sizes (from small well-coupled to large boulders/planetesimals), but suffers from Poisson noise, numerical diffusivity, computational overhead in grid interpolation, and inconsistency in the gas's back-reaction. A pressureless-fluid approach \citep[e.g.,][]{Zhu2012b,BenitezLlambay2019,Huang2022,Verrier2025,Ziampras2025b} alleviates these issues, but loses physical realism when dust trajectories are able to cross (typically, when the Stokes number $\St \equiv t_s \Omega_K \gtrsim 1$). Using multiple pressureless fluids to represent different size bins, \cite{Pfeil2024} and \cite{Fukuhara2025} compute a collisional $\beta$-cooling rate from the local dust abundance and grain-size distribution, albeit with a fixed background temperature and no live radiative transfer. In their radiation-hydrodynamics simulations, \cite{Binkert2021} and \cite{Krapp2024b} use local dust abundance and distribution to compute opacities in each cell, without considering collisional thermal relaxation.

A number of open questions in protoplanetary disk physics depend on the interplay of disk dynamics and thermodynamics simultaneously. For instance, the response of dust to spiral substructure is expected to vary as a function of grain size \citep[and therefore, collisional coupling time][]{Sturm2020,Speedie2022}, changing the observability of these substructures in the different wavelengths probing these grain sizes. Turbulence driven by disk instabilities, such as the vertical shear instability \citep[VSI; ][]{Nelson2013}, would impact the vertical distribution of dust \citep[e.g.,][]{Dullemond2022} and thereby the background temperature structure, which in turn would back-react onto the turbulent strength \citep[][]{MelonFuksman2023}. The puff-up of dust at the outer edge of a planet-driven gap \citep{Bi2021} could be expected to shadow the material behind it \citep{Dong2015c}, altering the disk's radial temperature profile and potentially giving rise to secondary instabilities. 

In order to investigate these outstanding progress, we have developed a numerical method to compute the dynamical and thermodynamical coupling between gas, radiation, and multiple species of dust. We implement this method within the grid-based, finite-volume hydrodynamics code PLUTO \citep{Mignone2007}. Section \ref{sec:intro} contains our introduction. Section \ref{sec:equations} presents and physically describes the equations solved by the method. Section \ref{sec:solution_strategy} describes the implicit-explicit (IMEX) scheme used to combine the implicitly integrated, potentially-stiff source terms with the non-stiff, explicit transport terms. Section \ref{sec:energy_momentum_exchange} presents the Newton-Raphson method used to iteratively solve the implicit terms; this is greatly accelerated by algebraic row-reduction of the Jacobian matrices for energy and momentum exchange, following  \cite{BenitezLlambay2019}. Section \ref{sec:advection_diffusion} presents the technique used to calculate the advection and diffusion of the dust. In Section \ref{sec:test_problems}, we present five test problems that verify the function and efficacy of our method, and in the following Section \ref{sec:sublimation}, we present a demonstrative application in the context of protoplanetary disks. Section \ref{sec:conclusion} summarizes our study and presents suggestions for future applications and numerical developments. Additional calculations are included in the Appendix.

\section{Equations}\label{sec:equations}
In what follows, we present the equations of radiation hydrodynamics with multiple fluids: a gas fluid with pressure, and one or more pressureless dust fluids. These fluids exchange energy and momentum with the radiation field via absorption and emission, and between gas and dust via collisional coupling. We evolve the gas energy using a total-energy scheme, whereas we treat dust energies as tracers advected along with the dust density. The radiation field is handled using a frequency-integrated moment method \citep{MelonFuksman2019,MelonFuksman2021} with the M1 closure, using the reduced speed of light approximation \citep[RSLA; ][]{Gnedin2001} to accelerate integration.

\begin{subequations}\label{eq:radhydro}
\begin{equation}
    \frac{\partial \rho_g}{\partial t} + \nabla \cdot (\rho_g \vec{v}_g) = 0
\end{equation}
\begin{equation}
    \frac{\partial \rho_{d,j}}{\partial t} + \nabla \cdot \left(\rho_d \vec{v}_{d,j}
    %) = -\nabla \cdot \left(
     + \vec{j}_{{\rm diff},j}\right) = 0
\end{equation}
\begin{equation}
    \frac{\partial (\rho_{g} \vec{v}_g)}{\partial t} + \nabla \cdot (\rho_g \vec{v}_g \vec{v}_g + \mathbf{I}p_g) = -\rho_{g} \nabla \Phi + \vec{S}_{m,g} + \vec{G}_{g} + \sum_{j=1}^{n_d} \vec{M}_{d,j}
\end{equation}
\begin{equation}
    \frac{\partial (\rho_{d, j} \vec{v}_{d,j})}{\partial t} + \nabla \cdot (\rho_{d,j} \vec{v}_{d,j} \vec{v}_{d,j} + \vec{j}_{{\rm diff},j} \vec{v}_{d,j}) = -\rho_{d,j} \nabla \Phi
    %+ \vec{S}_{m,d,j} 
    + \vec{G}_{d,j} - \vec{M}_{d,j}
\end{equation}
\begin{equation}
\begin{split}
    \frac{\partial \mathscr{E}_g}{\partial t} + \nabla \cdot (\mathscr{E}_g \vec{v}_g + (p + \rho_g \Phi) \vec{v}_g) = &\hphantom{+} S_{m,g} + \sum_{j=1}^{n_d} X_{gd, j} + cG_g\\
    &  + S^{\rm irr}_g + \vec{v}_g \cdot \sum_{j=1}^{n_d} \vec{M}_{d,j}  + Q_g
    \end{split}
\end{equation}
\begin{equation}
\label{eq:ed_evo}
    \frac{\partial E_{d,j}
    %+ K_{d,j})
    }{\partial t} + \nabla \cdot (E_{d,j}
    %+ K_{d,j}) 
    \vec{v}_{d,j} + (\vec{j}_{{\rm diff},j} /\rho_{d,j}) E_{d,j}) =
    -X_{gd,j} + cG_{d,j} + S^{\rm irr}_{d,j}% + S_{e, d, j}
\end{equation}
\begin{equation}
\label{eq:er_evo}
    \frac{\partial E_r}{\partial t} + \hat{c}\nabla \cdot \vec{F}_r = -\hat{c}\left(G_g + \sum_{j=1}^{n_d} G_{d,j}\right)
\end{equation}
\begin{equation}
\label{eq:flux_evo}
    \frac{\partial \vec{F}_r}{\partial t} + \hat{c}\nabla \cdot \mathbf{P}_r = -\hat{c}\left(\vec{G}_g + \sum_ {j=1}^{n_d}\vec{G}_{d,j}\right)
\end{equation}
\end{subequations}
In the above, $\rho$ represents density (either of the gas $g$, or species $j$ of dust $d$, as the subscript may indicate), $\vec{v}$ represents velocity, and $E$ represents internal energy. $\mathbf{I}$ is the identity matrix, $p$ the pressure, and $\mathscr{E}_g = E_g + 1/2 \rho_g v_g^2$ is the total (internal plus kinetic) energy of the gas. $\vec{S}_{m,g}$ describes the momentum source term arising from the divergence of the viscous stress tensor, $\mathbf{T} \equiv \nu\left[\nabla \vec{v} + (\nabla \vec{v})^{\top}\right]$, where $\nu$ denotes the kinematic viscosity; the scalar $S_{m,g}$ denotes viscous heating.

For the gas, we relate pressure, temperature, and energy using an ideal equation of state
\begin{equation}\label{eq:idealgas}
    E_g = \frac{p_g}{\gamma - 1} = \frac{1}{\gamma - 1}\frac{\rho_g k_B T_g}{\mu} \,,
\end{equation}
where $k_B$ is the Boltzmann constant, $T$ represents temperature (in this case, of the gas), $\mu$ the mean molecular weight, and $\gamma$ the adiabatic index.

\subsection{Radiation transport}
The radiation energy density $E_r$ and the radiative flux $\vec{F}_r$ are the zeroth and first moments of the radiative intensity, integrated over all frequency and solid angle. The evolution equations for the same (\ref{eq:er_evo} and \ref{eq:flux_evo}, respectively) are closed using the M1 approximation for the radiation pressure tensor
\begin{equation}
    \mathbf{P}_r \equiv E_r \left(\frac{1 - \Xi}{2} \mathbf{I} + \frac{3\Xi - 1}{2} \vec{n}\vec{n}\right) \,,
\end{equation}
in which $\vec{n} \equiv \vec{F}_r / ||\vec{F}_r||$, $w \equiv ||\vec{F}_r|| / E_r$, and 
\begin{equation}
    \Xi \equiv \frac{3 + 4w^2}{5 + 2 \sqrt{4 - 3w^2}} \,.
\end{equation}
This approximation reproduces both the optically thin free-streaming and optically thick diffusion limits of radiative transfer. In the evolution equations, $c$ is the speed of light, whereas $\hat{c}$ is the reduced speed of light; this change leads to a difference in the system's conservation laws, which we detail in Section \ref{sec:conservation_laws}.  We refer the reader to \cite{MelonFuksman2019} and \cite{MelonFuksman2021} for further details of the implementation.

\subsection{Radiative coupling}
In the rest frame of each of the respective species, the scalar source terms for energy exchange ($G$) and the vector source terms for momentum exchange ($\vec{G}$) with the radiation field are given by
\begin{equation}
\begin{split}
    G' &= \kappa \rho (E_r' - a_r T^4)\\
    \vec{G}' &= \chi \rho \vec{F_r}'
\end{split}
\,,
\end{equation}
where $T$ represents the temperature of the dust or gas species in question, $\rho$ its density, $\kappa$ its (frequency-integrated) absorption opacity, and $\chi \equiv \kappa + \sigma$ its total (absorption $\kappa$ plus scattering $\sigma$) opacity. All frequency-integrated opacities can, in general, be functions of any primitive variable (and by extension, temperature). To incorporate these terms into our method, we Lorentz-transform them into the laboratory frame, and use the procedure in \citep{MelonFuksman2021}---expanding to first order in $\vec{\beta} \equiv \vec{v}/c$, while keeping some second-order terms to enforce equilibrium:

\begin{equation}\label{eq:rel_terms}
\begin{split}
    G &= \rho\kappa \left(E_r - a_rT^4\right) - 2\kappa \rho \vec{\beta}\cdot \vec{F}_r + \rho \chi  \vec{\beta} \cdot \vec{F_r}\\
    & + \rho \chi \left( - E_r \vec{\beta} - \vec{\beta} \cdot \mathbf{P}_r\right) \vec{\beta}\\
    \vec{G} &= \rho \chi \vec{F_r} + \rho\kappa \left(E_r - a_rT^4\right)\vec{\beta} - 2\kappa \rho \vec{\beta}\cdot \vec{F}_r\vec{\beta}\\
    & + \rho \chi \left( - E_r \vec{\beta} - \vec{\beta} \cdot \mathbf{P}_r\right) \,. 
\end{split}
\end{equation}
Our formulation also admits the inclusion of an additional irradiation source term $S^{\rm irr}$ for each species, which is important when accounting for exogenous sources of heating such as stellar illumination (see Section \ref{sec:irradiation_test}). Although our method implements opacities and irradiation heating for the gas, we set their value to zero in the tests we present. 

\subsection{Collisional coupling}
Collisions between dust grains and gas particles cause momentum and energy to be exchanged between them. Collisional momentum exchange is given by \citep[e.g.,][]{BenitezLlambay2019}
\begin{equation}
    \vec{M}_{d,j} = \rho_{d,j} t_{d,j}^{-1}\left(\vec{v}_{d,j} - \vec{v}_g\right)
\end{equation}
while collisional energy exchange takes the form \citep[e.g.,][]{BurkeHollenbach83}
\begin{equation}
    X_{gd, j} = \eta_{d,j} \rho_{d,j} t_{d,j}^{-1} \left(2 k_B T_{d,j} - 2 k_B T_g\right)
\end{equation}

Within the quantity $\eta_{d,j} \equiv (3/4)\alpha_{\rm acc, d,j}/\mu$, $\mu$ is the mean molecular weight of the gas, $\alpha_{\rm acc, d,j}$ is an ``accommodation coefficient'' quantifying the efficiency of collisions at transferring energy, and the prefactor of $3/4$ arises from the assumption that grains are spherical. We note also $Q_{g}$, the frictional heating rate of the gas due to the dissipation of kinetic energy by $\vec{M}_{d,j}$, given by
\begin{equation}
    Q_g = \omega \sum_{j=1}^{n_d} \rho_{d,j}t_{d,j}^{-1} (\vec{v}_{d,j}-\vec{v}_g)^2 \,,
\end{equation}
where, following \cite{Huang2022}, $\omega$ is a parameter which can be set to 0 to turn frictional heating off, or 1 to turn it on. In general, frictional heating must be included for collisional coupling to conserve total energy (Section \ref{sec:conservation_laws}). We assume that all frictional heating is initially deposited in the gas (and subsequently shared with the dust collisionally), although more general formulations can be devised to distribute it as desired between all gas and dust species (Section \ref{sec:conclusion}).

The gas-grain stopping time $t_{d,j}$ is that required for the momentum of a dust particle to dissipate to gas molecules via collisional encounters. In general, our method allows $t_{d,j}$ to be any function of the fluid variables; for physically realistic expressions, valid in the subsonic regime ($\left|\vec{v}_{d,j} - \vec{v}_g\right| < c_s$), we refer the reader to previous literature \citep[e.g.,][]{Speedie2022,Muley2023}.

As with the stopping time, our solution method for energy coupling (Section \ref{sec:energy_momentum_exchange}) admits fully general expressions for dust energy and temperature. In practice, however, we compute gas temperatures from the pressure using the ideal equation of state in Equation \ref{eq:idealgas}. Likewise, for all tests we present in this work, we evaluate the dust temperature from the internal energy using the relation
\begin{equation}
    E_{d,j} = \rho_{d,j} c_{d,j} T_{d,j} \,,
\end{equation}
where $c_{d,j}$ is the specific heat capacity of the dust. This quantity is advected at the same velocity as the dust, and does not include its kinetic energy.

\subsection{Dust diffusion}
There are many possible functional forms for the dust diffusion flux $\vec{j}_{{\rm diff},j}$, but we opt for a gradient-diffusion prescription acting on the concentration %\citep[e.g.,][]{Weber2019}
\begin{equation}\label{eq:diffusion}
    \vec{j}_{{\rm diff},j} = -D \rho_g \nabla (\rho_{d,j} / \rho_g) = \rho_{d,j} \vec{v}_{{\rm diff},j} \,,
\end{equation}
where the diffusion coefficient is equal to the kinematic viscosity, which in our case is supplied by the $\alpha$-disk model \citep{Shakura1973}, $D = \alpha c_{s, \rm iso}^2 \Omega_{\rm turb}^{-1}$, where $c_{s, \rm iso} \equiv \sqrt{p_g/\rho_g} = \sqrt{k_B T_g/\mu}$ is the so-called ``isothermal sound speed'', and $\Omega_{\rm turb}$ is the typical turbulent-eddy turnover frequency, typically held to equal the Keplerian orbital frequency $\Omega_K$. The corresponding transport of dust momentum due to this concentration diffusion can be computed as
\begin{equation}\label{eq:sm_d_j}
    \vec{S}_{m,d,j} = -\nabla \cdot \left(\vec{j}_{{\rm diff},j} \vec{v}_{d,j}\right) = -\nabla \cdot \left(\rho_{d,j} \vec{v}_{{\rm diff},j} \vec{v}_{d,j}\right) \,.
\end{equation}
We can analogously define the scalar $S_{e, {\rm diff}, j}$, representing the transport of the dust internal energy:
\begin{equation}\label{eq:se_d_j}
    S_{e,d,j} = -\nabla \cdot \left((\vec{j}_{{\rm diff},j} /\rho_{d,j}) E_{d,j}\right) = -\nabla \cdot \left(E_{d,j} \vec{v}_{{\rm diff},j} \right) \,.
\end{equation}

Unlike more detailed formulations, such as those used in  \cite{Huang2022} and \cite{Binkert2023}, this simple prescription does not conserve the momentum generated by the diffusive flux itself, nor the angular momentum of the system overall. Nevertheless, we employ it for its simplicity of interpretation and implementation.
\section{Solution strategy}\label{sec:solution_strategy}
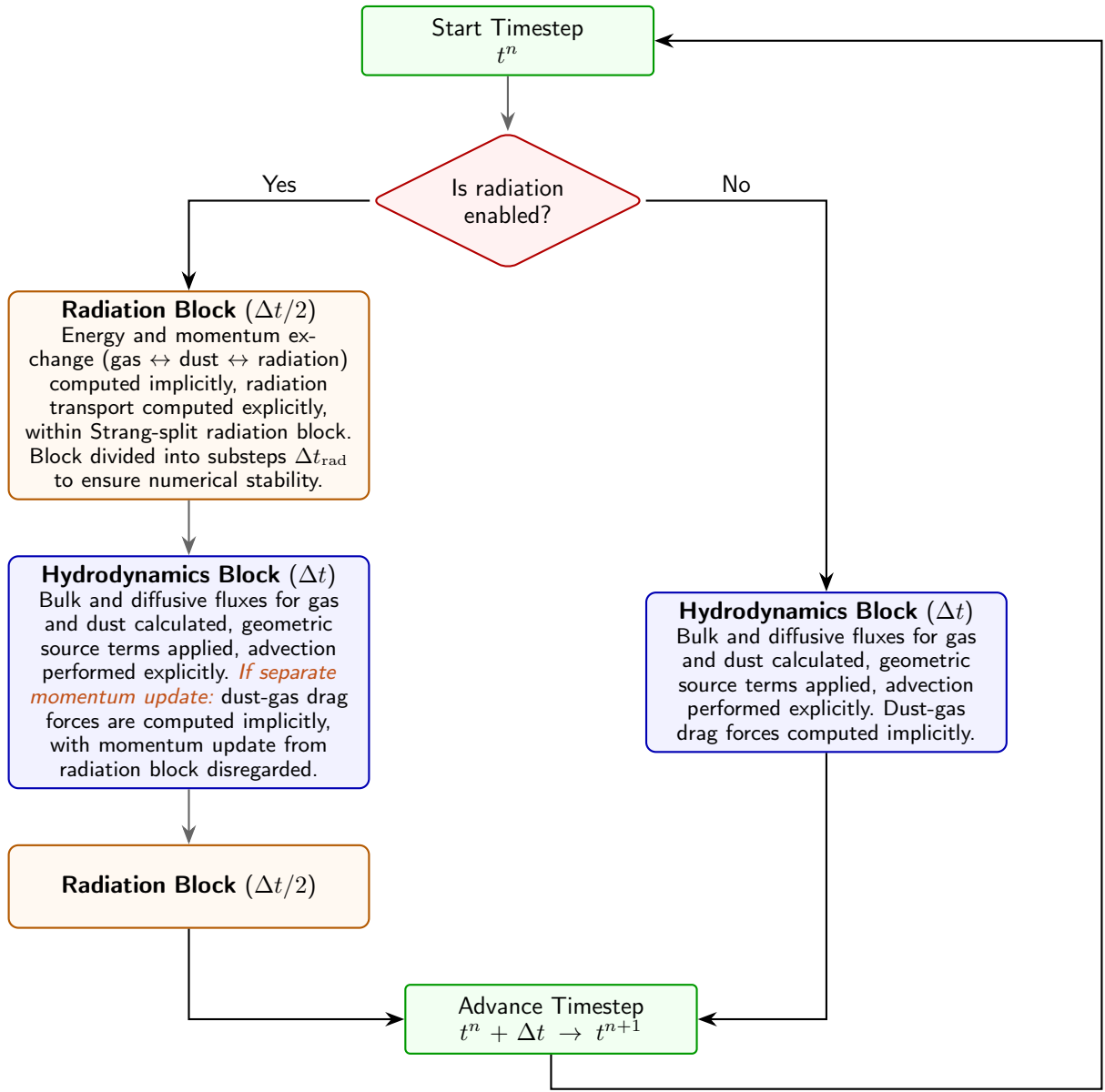
\begin{figure*}
    \centering
    \usetikzlibrary{shapes.geometric, arrows}
\usetikzlibrary{shapes.geometric, arrows.meta, positioning}
\tikzset{nodes={font=\sffamily}}

\begin{tikzpicture}[
    %- Setup block styles
    block/.style={
        rectangle, 
        draw=blue!70!black, 
        fill=blue!5, 
        thick, 
        text width=5cm, 
        align=center, 
        minimum height=1.2cm, 
        rounded corners
    },
    splitsource/.style={
        rectangle, 
        draw=orange!70!black, 
        fill=orange!5, 
        thick, 
        text width=5cm, 
        align=center, 
        minimum height=1.2cm, 
        rounded corners
    },
    decision/.style={
        diamond, 
        draw=red!70!black, 
        fill=red!5, 
        thick, 
        text width=2cm, 
        aspect=2,
        align=center, 
        minimum height=1.2cm, 
        rounded corners
    },
    loop/.style={
        rectangle, 
        draw=green!60!black, 
        fill=green!5, 
        thick, 
        text width=4cm, 
        align=center, 
        minimum height=1cm,
        rounded corners=2pt
    },
    arrow/.style={
        -{Stealth[scale=1.2]}, 
        thick, 
        draw=gray!80!black
    },
    line/.style = {draw, -{Stealth[scale=1.2]}, thick}
]

    %--- Nodes ---
    % Start
    \node [loop] (start) {Start Timestep \\ $t^n$};

    \node [decision, below=0.8cm of start] (rad_enabled) {Is radiation enabled?};

%RADIATION BRANCH:
    
    % Step 1: Radiation Half-Step
    \node [splitsource, below left=0.8cm and 1.0cm of rad_enabled] (rad1) {
        \textbf{Radiation Block} ($\Delta t / 2$) \\ 
        \vbox{\baselineskip=12pt \small Energy and momentum exchange (gas $\leftrightarrow$ dust $\leftrightarrow$ radiation) computed implicitly, radiation transport computed explicitly, within Strang-split radiation block. Block divided into substeps $\Delta t_{\rm rad}$ to ensure numerical stability.}
    };
    
    % Step 2: Hydro Half-Step
    \node [block, below=0.8cm of rad1] (hydro) {
        \textbf{Hydrodynamics Block} ($\Delta t$) \\ 
        \vbox{\baselineskip=12pt \small Bulk and diffusive fluxes for gas and dust calculated, geometric source terms applied, advection performed explicitly. \textit{\textcolor{Bittersweet}{If separate momentum update:}} dust-gas drag forces are computed implicitly, with momentum update from radiation block disregarded.}
    };
    
    % Step 3: Radiation Half-Step
    \node [splitsource, below=0.8cm of hydro] (rad2) {
        \textbf{Radiation Block} ($\Delta t / 2$) \\ 
        %\vbox{\baselineskip=12pt \small Energy and momentum exchange; radiation transport.}
    };

%NO RADIATION BRANCH:
    \node [block, right=4.0cm of hydro] (hydro_full) {
        \textbf{Hydrodynamics Block} ($\Delta t$) \\ 
        \vbox{\baselineskip=12pt \small Bulk and diffusive fluxes for gas and dust calculated, geometric source terms applied, advection performed explicitly. Dust-gas drag forces computed implicitly.}
    };

%separate energy and momentum coupling?
    %\node [decision, right=0.6cm of rad1] (sep_momentum) {Separate momentum?};
    
    % End of loop / Advance time
    \node [loop, below right=0.8cm and 0.5cm of rad2] (end) {Advance Timestep \\ $t^n + \Delta t \rightarrow t^{n+1}$};

    %--- Connections ---
    \draw [arrow] (start) -- (rad_enabled);

    \path [line] (rad_enabled) -| node[above, pos=0.25] {Yes} (rad1);
    \path [line] (rad_enabled) -| node[above, pos=0.25] {No} (hydro_full);
    
    %\draw [arrow] (rad_enabled) -- (rad1);
    \draw [arrow] (rad1) -- (hydro);
    \draw [arrow] (hydro) -- (rad2);
    %\draw [arrow] (rad2) -- (end);

    %\draw [arrow] (hydro_full) -- (end);
    \path [line] (rad2) |- (end.west);
    \path [line] (hydro_full) |- (end.east);

    %separate momentum
    %\path [line] (rad1.east) |- (sep_momentum.west);
    %\path [line] (sep_momentum) |- node[below, pos=0.25] {Yes} (hydro);
    %\path [line] (sep_momentum) |- node[below, pos=0.25] {No} (hydro_full);

    % Loop back to the start (routing around the right side)
    \path [line] (end.south) -- ++(0,-0.5) -- ++(8,0) |- (start.east);

    % Loop back arrow
    %\draw [arrow] (end.east) -- ++(1.5,0) |- (start.east) 
        node[pos=0.25, right, text=black] {Next Timestep};

\end{tikzpicture}
    \caption{A graphical summary of the solution strategies employed for this method. In simulations including radiation, the transport of radiation, as well as energy and momentum exchange between species, are handled within a radiation block Strang-split \citep{Strang1968} from the main hydrodynamical step. When radiation is disabled---or when radiation is enabled, but a separate momentum update is indicated---dust-gas momentum exchange is handled alongside the hydrodynamics.}

    \label{fig:solution_strategy}
\end{figure*}
We summarize the basic equations \ref{eq:radhydro} in the conservation-law form
\begin{equation}
    \frac{\partial\mathcal{U}}{\partial t} + \nabla \cdot \mathcal{F}(\mathcal{U}) = \mathcal{S}(\mathcal{U}) \,,
\end{equation}
where $\mathcal{U}$ is an array of conserved quantities, $\mathcal{F}$ a hyperbolic flux for these quantities, and  $\mathcal{S}$ the various right-hand-side source terms.

We present a broad outline of our solution strategy in Figure \ref{fig:solution_strategy}. The simulation timestep is determined by the Courant-Friedrichs-Lewy \citep[CFL;][]{CFL1928} criterion
\begin{equation}\label{eq:cfl_criterion}
    \Delta t = C_{\rm CFL} (\Delta x/v_{\rm sig})_{\rm min, grid} \,,
\end{equation}
where $\Delta x$ represents a cell size, $v_{\rm sig}$ the signal speed (including the diffusive velocity), and $C_{\rm CFL}$ a prefactor whose maximum permissible value depends on the time-integration scheme \citep{Gottlieb2001}; this value is computed for each grid cell, and the minimum over the grid taken as the timestep for all cells. 

For the radiation subsystem (Equations \ref{eq:er_evo} and \ref{eq:flux_evo}), the typical signal velocity ($v_{\rm sig, rad} \sim \hat{c}$) is much larger than that of the hydrodynamics ($v_{\rm sig, hydro} \sim c_s + |v_g|$). Given the timestep disparity, we follow \cite{MelonFuksman2021} and Strang-split each timestep, with each hydrodynamic block of duration $\Delta t$ sandwiched by two radiation blocks, as shown in the left case of Figure \ref{fig:solution_strategy}. Within the hydrodynamic block, we include advection, diffusion, pressure, viscous stresses, and body forces, as well as geometric source terms, for all gas and dust species as applicable. This block is integrated explicitly using a total-variation-diminishing, second-order Runge-Kutta (RK2) method \citep{Gottlieb1998}.

Each radiation block is divided into substeps of length $\Delta t_{\rm rad}$, determined by the radiation signal speed according to Equation \ref{eq:cfl_criterion}. Within each substep, the transport of radiation is integrated explicitly. Source terms for energy and momentum exchange between gas, dust, and radiation can, however, operate on even shorter timescales, so we integrate them implicitly to ensure numerical stability.\footnote{Following \cite{MelonFuksman2022}, the relativistic corrections in Equation \ref{eq:rel_terms}---typically of low magnitude in the context of protoplanetary disks---are integrated explicitly.} 

The final update of the conserved variables can then be computed using an implicit-explicit \citep[IMEX; see e.g.,][]{Pareschi2005} scheme. We opt for the IMEX1 method used by \cite{MelonFuksman2021} for its simplicity, stability, and ease of implementation:
    \begin{equation}\label{eq:imex1_method}
    \begin{split}
        \mathcal{U}^{(0)} - \mathcal{U}^n &= 0 \\
        \mathcal{U}^{(1)} -  \mathcal{U}^{(0)}&=  \Delta t (\nabla \cdot \mathcal{F} + \mathcal{S}_E)(\mathcal{U}^{(0)}) + \Delta t\mathcal{S}_I(\mathcal{U}^{(1)}) \\
        \mathcal{U}^{(2)} - \mathcal{U}^{(1)}  &= \Delta t (\nabla \cdot \mathcal{F} + \mathcal{S}_E)(\mathcal{U}^{(1)}) + \Delta t\mathcal{S}_I(\mathcal{U}^{(2)}) \\
        \mathcal{U}^{n+1} - \mathcal{U}^n &= \frac{\Delta t}{2}\left[(\nabla \cdot \mathcal{F} + \mathcal{S}_E)(\mathcal{U}^{(0)}) +  (\nabla \cdot \mathcal{F} + \mathcal{S}_E)(\mathcal{U}^{(1)})\right]\\
        &\hphantom{= } + \frac{\Delta t}{2}\left[\mathcal{S}_I(\mathcal{U}^{(1)}) + \mathcal{S}_I(\mathcal{U}^{(2)})\right]\\
         \rightarrow \mathcal{U}^{n+1} &= \frac{1}{2}\left[\mathcal{U}^{(2)} + \mathcal{U}^n \right] \,,
    \end{split}
    \end{equation}
where $\mathcal{S}_E$ represents the explicitly integrated source terms, and $\mathcal{S}_I$ the implicitly integrated source terms. In the limit that the implicit terms tend to zero, this method reproduces the RK2 method of \cite{Gottlieb1998}.

The radiation block represents a substantial computational expense, and may be replaceable with simplified thermodynamics treatments (e.g., $\beta$-cooling) in certain scientific applications. Therefore, we have also made it possible to run the module with gas and dust dynamics only. For these simulations (the no-radiation case of Figure \ref{fig:solution_strategy}), we integrate the hydrodynamics block using the IMEX1 method in Equation \ref{eq:imex1_method}, with dust-gas momentum coupling falling into $\mathcal{S}_I$. The separate dust energies, as well as the energy exchange between gas and dust, are not accounted for in such integrations.

In some situations, such as when dust experiences inward radial drift due to gas headwinds (see e.g. Sections \ref{sec:radial_drift}), velocity is set by a dynamic equilibrium between explicitly (pressure, gravity, geometric) and implicitly integrated (dust-gas drag) source terms. In fully radiative runs employing solution strategy (a), we find that the operator split between the explicit and implicit terms introduces splitting error in the velocity, which can become substantial in the regime that $t_{d,j} \ll \Delta t$---precisely that in which an implicit method is useful. For this reason we also implement a hybrid scheme in which thermodynamic coupling is computed within the radiation block, but dynamical coupling is computed within the hydrodynamic block, as in the unsplit strategy. Doing so yields more accurate dynamic-equilibrium velocities, at the cost of error in frictional heating and radiant pressure---both of which are typically minimal in the context of protoplanetary disks.

\begin{comment}
\dmnote{Some bullets for this section:
\begin{enumerate}

\item In fully radiative runs, the fact that dust-gas drag is located in the radiation block, and thus operator-split from various other transport-related source terms (pressure, gravity, geometric, etc.) introduces error in the velocities/momenta at the end of the first Runge-Kutta phase $\mathcal{U}^{(1)}$, and in turn, in mass transport during the second phase $\mathcal{U}^{(2)}$.

as if dust-gas drag were completely absent. To remedy this issue, during the hydrodynamic computation, we implicitly compute dust-gas velocity coupling during the first stage of the RK2 integration, but then subtract it off of the final stage. (EDIT: we could even just throw away the information about momentum coupling in radiation, and apply IMEX1 during the hydro step. Or something more creative...)
\end{enumerate}}

\dmnote{We need some special techniques to ensure total local energy conservation here. Nevertheless, this cannot conserve energy globally because there is no P dV term for dust, i. e. no dust pressure which could be set up in an adiabatic way to conserve the energy in converging dust flows. Dust pressure is covered in certain works (Klahr and Schreiber, Binkert 2023) and we believe it would be a useful future investigation after this work.}
\end{comment}
\section{Energy and momentum exchange}\label{sec:energy_momentum_exchange}

\subsection{Iterative procedure}\label{sec:iterative_procedure}
We use a Newton-Raphson technique to carry out the implicit update of the energy $\vec{E}^{i} \rightarrow \vec{E}^{i+1}$ and momenta $\vec{p}^{i} \rightarrow \vec{p}^{i+1}$. At each iteration $k$, we insert an estimate for the energy at that iteration, $\vec{E}^{i+1,*k}$, into the original equations (Equation \ref{eq:energy_update} for the energy, and Equation \ref{eq:momentum_exchange} for momentum) in order to compute the residuals $\vec{X}^{*k}$. To iterate the energy, we solve the system
\begin{equation}
    \delta \vec{X}^{k} - \mathbf{\tilde{J}} \delta \vec{E}^{i+1, *k} = 0 \,,
\end{equation}
where $\mathbf{\tilde{J}}$ represents the Jacobian matrix of the system, and where we identify $\delta \vec{X}^{*k} = -\vec{X}^{*k}$ and $\delta \vec{E}^{i+1, *k} = \vec{E}^{i+1,*k+1} - \vec{E}^{i+1, *k}$. This procedure is then repeated until each element of $\vec{X}^{*k}$ equals zero within a specified tolerance, or after a certain number of iterations is reached. Momentum is updated using an analogous method.

Both the energy and momentum updates require solving $(n_d + 2)$ equations dependent on $(n_d + 2)$ unknowns. However, the dust species do not directly interact with one another, and can only influence one another through their effects on gas and dust energy and momenta. This makes the matrix sparse and amenable to row-reduction, yielding a $2\times 2$ reduced Jacobian for energy coupling and a $1 \times 1$ for momentum coupling. These reduced systems are then used to update all of the gas energy and momenta one by one. This technique---analogous to that deployed by \cite{BenitezLlambay2019} and \cite{Krapp2024} for the momentum-coupling problem---reduces the computational complexity of the problem from $\mathscr{O}((n_d + 2)^3)$ to $\mathscr{O}(n_d)$, a significant benefit given that iterations are carried out multiple times per timestep.
\subsection{Energy evolution}\label{sec:energy_evolution}
We solve the following set of implicit equations, in which $\rho$ denotes density, $\kappa$ denotes Planck-averaged opacity, $T$ denotes temperature, and $E$ denotes internal energy for species $j$ (with subscript $r$ denoting radiation and $g$ denoting gas). $i$ denotes the value of a quantity at the start of an implicit partial step $\delta t$, and $i + 1$ denotes that at the end:
\begin{subequations}\label{eq:energy_update}

\begin{equation}
\begin{split}
    \frac{E_r^{i+1} - E_r^i + X_r}{\delta t}
    &={}
    -\hat{c}\,\kappa_g^{i+1}\rho_g
    \left(E_r^{i+1} - a_r T_g^{4,i+1}\right)
    \\
    &-\sum_{j=1}^{n_d}
    \hat{c}\,\kappa_{d,j}^{i+1}\rho_{d,j}
    \left(E_r^{i+1} - a_r T_{d,j}^{4,i+1}\right)
\end{split}
\end{equation}

\begin{equation}
\begin{split}
    \frac{E_g^{i+1} - E_g^i + X_g}{\delta t}
    &={}
    c\,\kappa_g^{i+1}\rho_g
    \left(E_r^{i+1} - a_r T_g^{4,i+1}\right)
    \\
    &+\sum_{j=1}^{n_d}
    \eta_{d,j}^{i+1}\rho_{d,j}t_{d,j}^{-1,i+1}
    \left(
        2k_B T_{d,j}^{i+1}
        -2k_B T_g^{i+1}
    \right)
    \\
    &+S_g+Q_g^{i+1}
\end{split}
\end{equation}

\begin{equation}
\begin{split}
    \frac{E_{d,j}^{i+1} - E_{d,j}^i + X_{d,j}}{\delta t}
    &={}
    c\,\kappa_{d,j}^{i+1}\rho_{d,j}
    \left(E_r^{i+1} - a_r T_{d,j}^{4,i+1}\right)
    \\
    &-\eta_{d,j}^{i+1}\rho_{d,j}t_{d,j}^{-1,i+1}
    \left(
        2k_B T_{d,j}^{i+1}
        -2k_B T_g^{i+1}
    \right)
    \\
    &+S_{d,j}
    \qquad \{j\geq 1\}
\end{split}
\end{equation}
\end{subequations}

We present the Jacobian matrices associated with these equations in Section \ref{sec:rad_matrix_elements} of the Appendix.

\subsection{Momentum evolution}
The implicit momentum update is carried out in a similar way to that of the energy, according to the following equations:
\begin{subequations}\label{eq:momentum_exchange}
\begin{equation}
    \frac{\vec{F}_r^{i+1} - \vec{F}_r^{i} + \vec{R}_r}{\delta t} = -\hat{c}\left(\rho_g \chi_g^{i+1} + \sum_{j=1}^{n_d}\rho_{d,j}\chi_{d,j}^{i+1}\right) \vec{F}_r^{i+1} \,,
\end{equation}
\begin{equation}
    \frac{\vec{p}_g^{i+1} - \vec{p}_g^i + \vec{R}_g}{\delta t} = \rho_g\chi_g^{i+1}\vec{F}_r^{i+1} + \sum_{j=1}^{n_d} \rho_{d,j}t_{d,j}^{-1,i+1}\left(\vec{v}_{d,j}^{i+1} - \vec{v}_{g}^{i+1}\right) \,,
\end{equation}
\begin{equation}
    \frac{\vec{p}_{d,j}^{i+1} - \vec{p}_{d,j}^i + \vec{R}_{d,j}}{\delta t} = \rho_{d,j}\chi_{d,j}^{i+1}\vec{F}_r^{i+1} - \rho_{d,j}t_{d,j}^{-1,i+1}\left(\vec{v}_{d,j}^{i+1} - \vec{v}_{g}^{i+1}\right) \,,
\end{equation}
\end{subequations}
where $\vec{R}$ is the vector of Newton-Raphson residuals for the momentum update. The linearization of the above, used to perform the Newton-Raphson iterations, is given in \ref{sec:momentum_matrix_elements}. The momentum update results in a net change in kinetic energy
\begin{equation}
\begin{split}
    \Delta K^{i+1} = &\frac{(\vec{p}_g^{i+1} - \vec{p}_g^i) \cdot (\vec{p}_g^{i+1} + \vec{p}_g^i)}{2\rho_g}\\
     &+ \sum_{j=1}^{n_d} \frac{(\vec{p}_{d,j}^{i+1} - \vec{p}_{d,j}^i) \cdot (\vec{p}_{d,j}^{i+1} + \vec{p}_{d,j}^i)}{2\rho_{d,j}} \,,
\end{split}
\end{equation}

of which an amount
\begin{equation}
    Q_r^{i+1} = \vec{F}_r^{i+1} \cdot \left[\chi_g \frac{\vec{p}_g^{i+1} + \vec{p}_g^{i}}{2} + \sum_{j=1}^{n_d} \chi_{d,j} \frac{\vec{p}_{d,j}^{i+1} + \vec{p}_{d,j}^{i}}{2} \right]
\end{equation}

represents conversion of radiation to kinetic energy (of all gas and dust species), and the difference $Q_g^{i+1} = \omega (\Delta K^{i+1} - Q_r^{i+1})$ represents frictional losses. Because $Q_r^{i+1}$ is already represented by the explicitly-integrated relativistic terms in Equation \ref{eq:rel_terms} (namely, those of the form $\rho \chi \vec{F_r} \cdot \vec{\beta}$ for each gas and dust species), we subtract it from the implicit step to avoid double-counting. Following \cite{Huang2022}, we apply the entirety of the frictional heating $Q_g^{i+1}$ to the gas.

\subsection{Conservation laws}\label{sec:conservation_laws}
Up to frictional and irradiation source terms, the implicit step conserves the sum of all internal energies and momenta. The use of the RSLA means that, in this sum, the radiation energy $E_r$ and the radiative flux $\vec{F}_r$ are weighted by factors of $(c/\hat{c})$ and $(1/\hat{c})$ respectively:
\begin{subequations}
\begin{equation}
    E_{\rm int, tot}^{i+1} - E_{\rm int, tot}^{i} = \left(Q_g^{i+1} + S^{\rm irr}_{g} + \sum_{j = 1}^{n_d} S^{\rm irr}_{d,j}\right)\delta t
\end{equation}
\begin{equation}
    \vec{p}_{\rm tot}^{i+1} - \vec{p}_{\rm tot}^{i} = \vec{0}
\end{equation}
\end{subequations}
where we define
\begin{subequations}
\begin{equation}
    E_{\rm int, tot} = \frac{c}{\hat{c}}E_r + E_g + \sum_{j=1}^{n_d} E_{d,j}\,,
\end{equation}
\begin{equation}\label{eq:mom_total_def}
    \vec{p}_{\rm tot} = \frac{1}{\hat{c}}\vec{F}_r + \rho_g \vec{v}_g + \sum_{j=1}^{n_d} \rho_{d,j} \vec{v}_{d,j}\,.
\end{equation}
\end{subequations}
The conservation properties of the system are demonstrated in Appendix \ref{sec:conservation_laws_appendix}. When frictional heating is on ($\omega = 1$), it can also be shown that the total energy $E_{\rm tot}$ also obeys the following conservation law during the implicit step:
\begin{equation}
    E_{\rm tot}^{i+1} - E_{\rm tot}^{i} = \left(S^{\rm irr}_{g} + \sum_{j = 1}^{n_d} S^{\rm irr}_{d,j}\right)\delta t \,,
\end{equation}
where
\begin{equation}\label{eq:energy_total_def}
    E_{\rm tot} = \frac{c}{\hat{c}}E_r + \mathscr{E}_g + \sum_{j=1}^{n_d} E_{d,j} + \sum_{j=1}^{n_d} \frac{1}{2}\rho_{d,j} \vec{v}_{d,j}^2 \,.
\end{equation}
In other words, total energy within the system remains constant, modified only by the injection of energy to the system via stellar irradiation.

\begin{comment}
A different addition of the linearized equations yields
\begin{equation}\label{eq:fast_linear_2}
    \delta \vec{R}_g + \sum_{j=1}^{n_d} \delta \vec{R}_{d,j} = -\delta \vec{p}_g^{i+1} - \sum_{j=1}^{n_d} \delta \vec{p}_{d,j}
\end{equation}
which, when subtracted from the original \ref{eq:fast_linear}, yields
\begin{equation}\label{eq:fast_linear_sum}
    -\sum_{j=1}^{n_d} \delta \vec{R}_{d,j} \frac{1}{1 + t_{d,j}^{-1,i+1} \delta t} = -\delta \vec{p}_g^{i+1}\left[ \sum_{j=1}^{n_d}   \frac{\rho_{d,j}}{\rho_g} \left(\frac{t_{d,j}^{-1,i+1} \delta t}{1 + t_{d,j}^{-1,i+1} \delta t}\right)\right] - \sum_{j=1}^{n_d}\delta \vec{p}_{d,j}^{i+1}
\end{equation}
\end{comment}

\section{Advection and diffusion}\label{sec:advection_diffusion}
Building on the dust-fluid module introduced for PLUTO by \cite{Ziampras2025b}, we evaluate dust advection and diffusion by assuming each dust species to be a pressureless fluid, and then solving a Riemann problem at cell interfaces. For the left and right states of dust density, internal energy, and bulk velocity, we use the standard, slope-limited reconstructions already present in \texttt{PLUTO}. For the diffusion velocity, we rearrange and discretize Equation \ref{eq:diffusion} to compute a first-order face-centered difference, yielding the following functional form in the simple case of the $x$-direction on a Cartesian grid:
\begin{equation}
    (\vec{v}_{\rm diff,j})_{x, i+1/2} = -\nu_{i+1/2} \frac{\left(\ln(\rho_{d,j}/\rho_g)_{i+1} - \ln(\rho_{d,j}/\rho_g)_{i}\right)}{(\Delta x_i/2 + \Delta x_{i+1}/2) }
\end{equation}
in which $x_{i+1/2}$ is the interface between cells $x_i$ and $x_i+1$. The viscosity is computed using the full set of primitive variables reconstructed at the left state $L$ or right state $R$, depending on whether the left or right state of the diffusion velocity is sought. In practice, we find that large mass gradients (many orders of magnitude across one cell) lead to crashes, so we limit the diffusion velocity to some
\begin{equation}
    v_{\rm diff, lim} = C_{\nu} \sqrt{\nu/t_{\nu}}
\end{equation}
where $C_{\nu}$ is a numerical constant and $t_{\nu}$ represents a characteristic timescale for diffusion. In protoplanetary disks, where dust diffusion is driven by turbulent eddies and $t_{\nu}$ can be set to the eddy turnover timescale (typically one dynamical time, $\Omega_K^{-1}$), this formulation of the limiter has a natural physical interpretation: $C_{\nu}$ represents the maximum logarithmic concentration difference that can be sustained across the lengthscale of a single turbulent eddy. Our tests reveal that $C_{\nu} \sim 100$ is sufficient to ensure stability.

The fluxes between cells can be computed as

\begin{align}
    (\mathbf{F}_{d,j})_{x,i+1/2} &= \begin{bmatrix}
           (\rho_{d,j})_{i+1/2} \left[\vec{v}_{d,j,x}  +\vec{v}_{\rm diff,j,x}\right]_{i+1/2} \\
            (\rho_{d,j} \vec{v}_{d,j,x})_{i+1/2} \left[\vec{v}_{d,j,x} + \vec{v}_{\rm diff,j,x}\right]_{i+1/2} \\
            (\rho_{d,j} \vec{v}_{d,j,y})_{i+1/2} \left[\vec{v}_{d,j,x} + \vec{v}_{\rm diff,j,x}\right]_{i+1/2} \\
            (\rho_{d,j} \vec{v}_{d,j,z})_{i+1/2} \left[\vec{v}_{d,j,x} + \vec{v}_{\rm diff,j,x}\right]_{i+1/2} \\
           (E_{d,j})_{i+1/2} \left[\vec{v}_{d,j,x} + \vec{v}_{\rm diff,j,x}\right]_{i+1/2}
         \end{bmatrix}
  \end{align}
which, again, can be evaluated at either the left and right states at the interface. The bulk velocity $\vec{v}_{d,j}$ is obtained through standard reconstruction techniques within PLUTO, such as the piecewise linear (PLM) or piecewise parabolic (PPM) methods. We compute the actual flux at the interface using the \cite{LeVeque2003} Riemann solver for a pressureless fluid, already used by \cite{Ziampras2025b}. For this solver, the sole (triply degenerate) characteristic velocity is
\begin{equation}
    \lambda_x = \frac{\sqrt{\rho_{d,j, L}} \left[\vec{v}_{d,j,x}  + \vec{v}_{\rm diff,j,x}\right]_{L} + \sqrt{\rho_{d,j, R}}\left[\vec{v}_{d,j,x}  + \vec{v}_{\rm diff,j,x}\right]_{R}}{\sqrt{\rho_{d,j, R}} + \sqrt{\rho_{d,j, L}}}
\end{equation}
For $\lambda_x < 0$, we assume the interface flux $(\mathbf{F}_{d,j})_* = (\mathbf{F}_{d,j})_R$, and for $\lambda_x > 0$, we set $(\mathbf{F}_{d,j})_* = (\mathbf{F}_{d,j})_L$. In the edge case that $\lambda_x = 0$, we take $(\mathbf{F}_{d,j})_* = (1/2)((\mathbf{F}_{d,j})_L + (\mathbf{F}_{d,j})_R)$. 

\section{Test problems}\label{sec:test_problems}
\subsection{Energy and momentum coupling}\label{sec:coupling_test}
To verify the basic functionality of our method, we first consider a 1D, Cartesian, periodic box over a range $0 \leq x < 1$, with a resolution of 800 cells in the $x$-direction. We simulate one gas species $g$ and five dust species, $1 \leq j \leq 5$, and assign the initial conditions as described in \Cref{tab:initial_conditions}; in this table, we assume that $\rho_0 = \SI{7.78e-10}{\gram\per\cm\tothe{3}}$, $\kappa_0 \equiv 2.17 \times 10^2$, and $t_0 = 8$ d. We assume an adiabatic index $\gamma \equiv 1.4$, $c = \SI{299792.458}{\meter\per\second}$, and $\hat{c} = 10^{-3}c$. For all dust species, we set the opacity $\kappa_{d,j} = \kappa_0$; moreover, we set the dust heat capacity equal to a factor $\gamma - 1$ that of the gas heat capacity, $c_g = k_B/(\gamma - 1)$. We use HLLC Riemann solvers for the transport of gas and radiation; we use the previously-mentioned LeVeque solver for dust, with the viscous diffusion coefficient $\nu$ set to zero.

\begin{table}
\centering
\begin{tabular}{|l|l|}
\hline
Variable & Value \\
\hline
    $\rho_g$       & $10^{-2}\rho_0$ \\
    $\rho_{d,j}$   & $10^{-5}\rho_0$ \\
    $v_{g,x}$      & $-6~\mathrm{km\,s^{-1}}$ \\
    $v_{d,j,x}$    & $10^{-2}\times0.25^{(j-1)}v_{g,x}$ \\
    $t_{d,j}$      & $10^{-5}\times0.5^{(j-1)}t_0$ \\
    $T_g$          & $10~\mathrm{K}$ \\
    $T_{d,j}$      & $100~\mathrm{K}\times0.5^{(j-1)}$ \\
    $T_r$          & $100~\mathrm{K}$ \\
    $F_{r,x}$      & $0$\\
    \hline
\end{tabular}
\\
\caption{Initial conditions used for the energy-momentum coupling test in \Cref{sec:coupling_test}.}
    \label{tab:initial_conditions}
\end{table}

\begin{figure}
    \centering
    \includegraphics[width=1.0\linewidth]{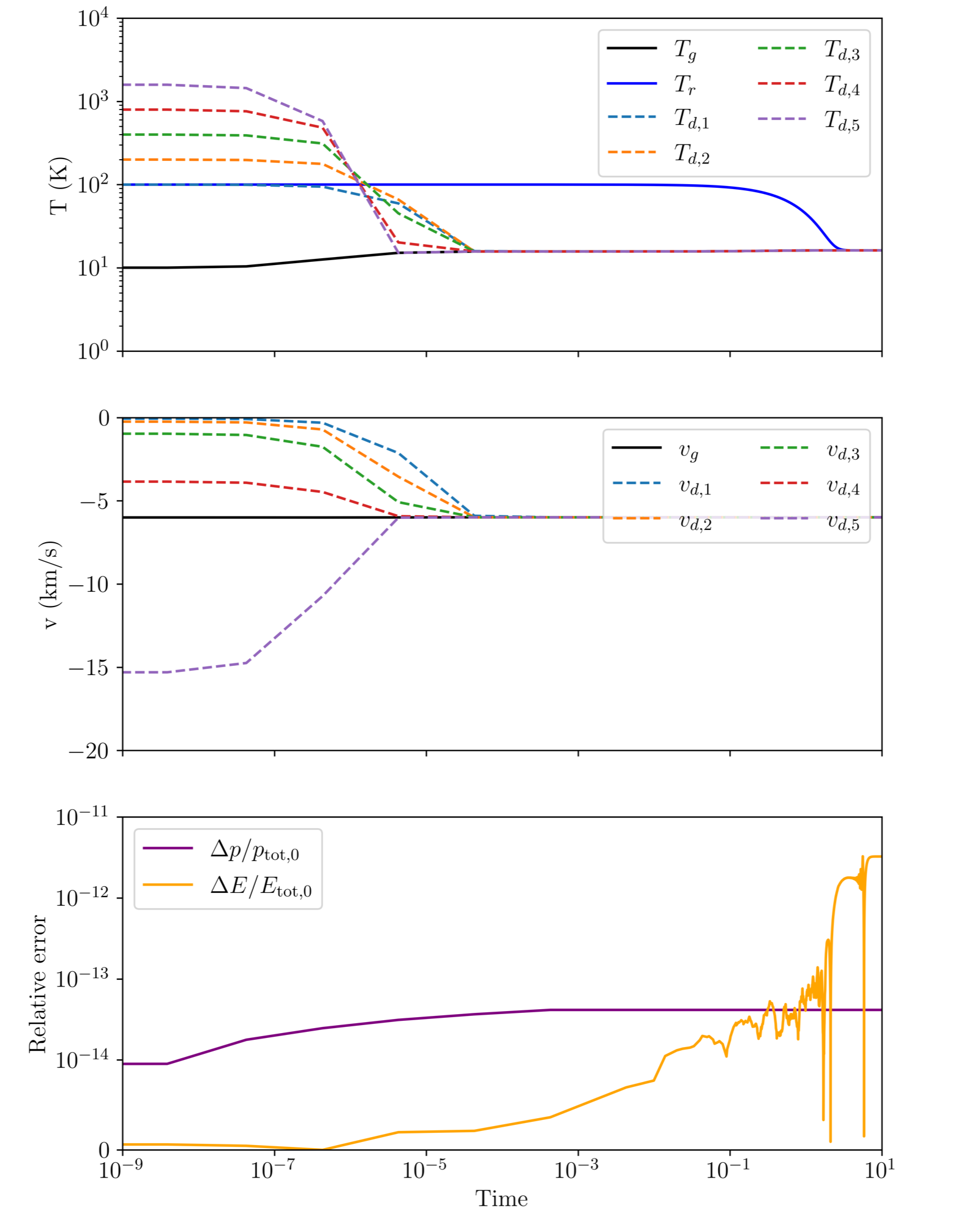}
    \caption{Relaxation of temperature, in the upper panel, and velocity, in the middle panel, for gas, radiation, and all dust species. In the lower panel, we plot the error in both total energy and total momentum. By the time the simulation has reached equilibrium, the energy error $\Delta E/E_0 \lesssim 10^{-11}$ and the momentum error $\Delta p/p_0 \lesssim 10^{-13}$, demonstrating the suitability of our scheme for long-term integrations.}
    \label{fig:temp_mom_relaxation}
\end{figure}

In the absence of any sources, sinks, or gradients in energy or material density, the only processes governing the dynamics and thermodynamics of the system should be the energy and momentum exchange mechanisms we have presented. In Figure \ref{fig:temp_mom_relaxation}, we plot the evolution of the temperatures and velocities of all species in the top and middle panels, respectively. Over the lifetime of the simulation, all species relax to the same temperature and velocity, at which point the energy and momentum exchange terms net to zero and the system is in equilibrium. Owing to the very short stopping times we use, the dust and gas rapidly equilibrate with one another, whereupon the dust-gas mixture slowly relaxes to a common temperature with the radiation. As with the 0D test in \cite{Muley2023}, this result demonstrates that our scheme is capable of reproducing the ``two-temperature'' limit used in most radiation-hydrodynamics schemes \citep[e.g.,][]{Kolb2013,MelonFuksman2021}, in the case that dust and gas are tightly coupled to one another. 

In the lower panel of Figure \ref{fig:temp_mom_relaxation}, we plot the change in total energy and momentum in the system over time, relative to the initial condition. The total energy and momentum are computed in accordance with Equations \ref{eq:energy_total_def} and \ref{eq:mom_total_def}, respectively.
\begin{comment}
For this calculation, we compute the total energy and momentum as
\begin{subequations}
\begin{equation}
    E_{\rm tot} = \frac{c}{\hat{c}}E_r + \mathscr{E}_g + \sum_{j=1}^{n_d} E_{d,j} + \sum_{j=1}^{n_d} \frac{1}{2}\rho_{d,j} \vec{v}_{d,j}^2
\end{equation}
\begin{equation}
    \vec{p}_{\rm tot} = \frac{1}{\hat{c}}\vec{F}_r + \rho_g \vec{v}_g + \sum_{j=1}^{n_d} \rho_{d,j} \vec{v}_{d,j}
\end{equation}
\end{subequations}
\end{comment}
The manifest conservativeness of the exchange terms, and the fact that our setup is a closed system, imply that the deviation in energy and momentum should be zero. At the implementation level, however, floating-point errors lead to small departures from true conservation ($\Delta E/E_{\rm tot, 0} \lesssim 10^{-11}$, $\Delta p/p_{\rm tot, 0} \lesssim 10^{-13}$), which stop growing once equilibrium has been reached. 

In addition to equilibrium and conservative properties, it is useful to verify that the scheme accurately reproduces the expected time evolution for temperature. In this test, the gas-dust collisional timescale is much shorter than the raditive timescale, meaning that at early times, the evolution of the system is dominated by thermal accommodation and obeys the following equation:

\begin{equation}
    \frac{d\vec{E}}{dt} = \tilde{\mathbf{X}}\vec{E} + \vec{Q}(t)  \,,
\end{equation}

\begin{comment}
\begin{equation}
\tilde{\mathbf{X}} = \begin{bmatrix}
0 & 0 & 0 & 0 & ...\\
0 & \sum_{j=1}^{n_d} \eta_{d,j} \rho_{d,j} t_{d,j}^{-1} & -\eta_{d,1} \rho_{d,1} t_{d,1}^{-1} & -\eta_{d,2} \rho_{d,2} t_{d,2}^{-1} & ... \\
0 & -\eta_{d,1} \rho_{d,1} t_{d,1}^{-1} & \eta_{d,1} \rho_{d,1} t_{d,1}^{-1} & 0 & ...\\
0 & -\eta_{d,2} \rho_{d,2} t_{d,2}^{-1} & 0 & \eta_{d,2} \rho_{d,2} t_{d,2}^{-1} & ...\\
... & ... & ... & ...& ...\\
\end{bmatrix}
\end{equation}
\end{comment}

where $\vec{E}$ represents a list of energies, $\tilde{\mathbf{X}}$ the matrix associated with collisional energy exchange, and $\vec{Q} = \left(0, Q_g, 0, 0,...\right)^\top$ represents the frictional heating term. This system can be solved using the method of integrating factors to yield that
\begin{equation}\label{eq:analytical_solution_energy}
    \vec{E}(t) = \exp(\tilde{\mathbf{X}}t) \vec{E}_0 + 
    \int_0^t \exp(-\tilde{\mathbf{X}}(t
    ' - t)) \vec{Q}(t') dt' \,.
\end{equation}

\begin{figure}
    \centering
    \includegraphics[width=1.0\linewidth]{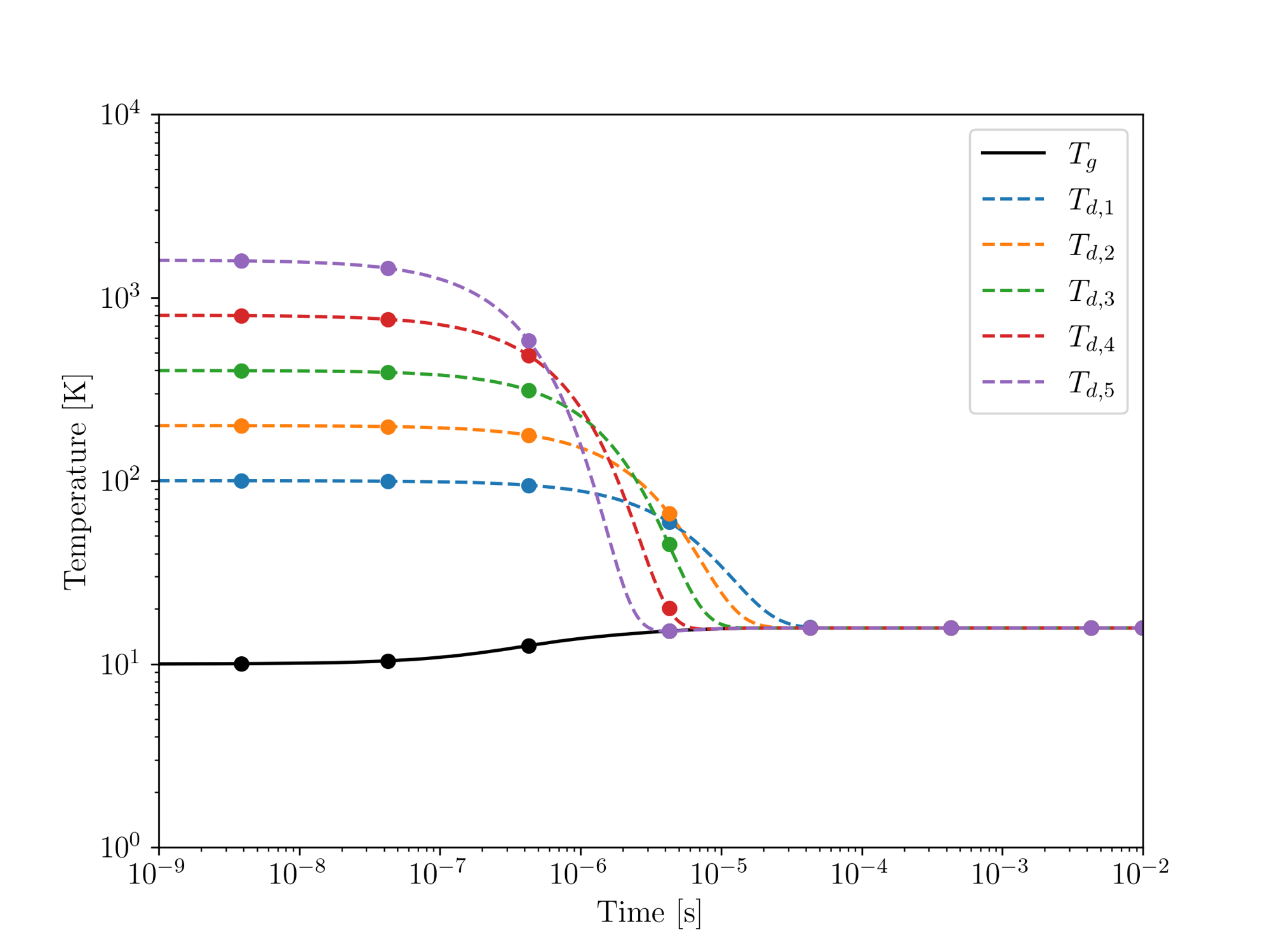}
    \caption{Dust and gas temperatures at early times in the coupling times, computed numerically (points) or semi-analytically using Equation \ref{eq:analytical_solution_energy} (curves). The simulated temperatures show good agreement with the semi-analytic estimates, demonstrating simultaneously the correct implementation of collisional energy exchange, collisional momentum exchange, and frictional heating.}
    \label{fig:analytical_vs_numerical_0d}
\end{figure}

The homogeneous term, representing exchange of the pre-existing thermal energy, can be solved analytically. The integrand of the inhomogeneous term can be computed using the fact that $Q_g \equiv \sum_{j=1}^n \rho_{d,j} t_{d,j}^{-1} (\vec{v}_{d,j} - \vec{v}_g)^2$, in which the velocities can be derived from the analytical evolution of the momenta:
\begin{equation}
    \vec{p}(t) = \exp(\mathbf{\tilde{M}}t) \vec{p}_0 \,, 
\end{equation}
where $\mathbf{\tilde{M}}$ is the matrix containing momentum-exchange terms. The integral itself, however, is carried out using the trapezoidal method. In Figure \ref{fig:analytical_vs_numerical_0d}, we plot this semi-analytical expectation against the numerical solution obtained from the multispecies method. The two show close agreement, demonstrating that the method correctly implements not just energy exchange, but also the momentum exchange on which the value of frictional heating depends.

\subsection{Ray-traced irradiation}\label{sec:irradiation_test}
In addition to the internal energy-exchange mechanisms studied in the previous Section \ref{sec:coupling_test}, various external heating and cooling processes---such as ray-traced starlight \citep[e.g.,][]{MelonFuksman2022} or planetary accretion luminosity \citep{Muley2024}---play a crucial role in determining the thermal properties of protoplanetary disks. In numerical simulations, such processes are often represented using an irradiation source term $S$. Our work generalizes these previous ones by including separate source terms for gas ($S_g$) and each of the various dust species ($S_{d,j}$)---a particularly relevant feature for multi-frequency, ray-traced stellar illumination, due to the varying opacities that grains of different sizes and compositions have in the various frequency bands constituting a stellar spectrum.

We use the setup from the previous section as a starting point, assuming the same constant dust-to-gas ratio ($f_{d,j} = 10^{-3}$) and thermal-infrared absorption opacity ($\kappa_{d,j} = \kappa_{0}$) for each of the various dust species. This time, we set the initial temperatures of all species to 10 K, and their initial velocities to zero. For the irradiation flux, we assign each dust species its own, frequency-dependent absorption opacity
\begin{equation}
    (\kappa_{d, \rm irr})_{j,b} = 10^4 \kappa_0 (j)^{-1}(b)^{-1}
\end{equation}
where as before $j$ is the index of the dust, and $b$ represents the index of the given frequency band, starting from 1. In this test, we use 5 dust species and 3 frequency bands, with no gas opacity either in the thermal infrared or for irradiation.

At the left edge of the domain, we initialize a total flux as a sum over contributions from individual frequency bands, $\vec{F}_0 \equiv \sum_b \vec{F}_{0,b}$, with rays traced from left to right:
\begin{equation}\label{eq:irr_flux}
\begin{split}
    \vec{F}(x) &= \sum_{b=1}^{b_{\rm max}} \vec{F}_{b}(x)\\
     &= \sum_{b=1}^{b_{\rm max} }\vec{F}_{0,b}(x) \exp\left(- \int_0^x\sum_{l=1}^{n_d} \rho_{d,l} (\kappa_{d, \rm irr})_{l,b} dx' \right) \,.
\end{split}
\end{equation}
The irradiation heating such a flux produces in a given cell $i$ is:
\begin{equation}\label{eq:irr_term_exp}
    \left<S_{\rm irr,d,j}\right>_i = \sum_{b = 1}^{b_{\rm max}} \frac{c \kappa_{d,j,b} \rho_{d,j}}{\sum_{l=1}^{n_d}\kappa_{d,l,b} \rho_{d,l}} \left<-\nabla \cdot \vec{F}_{b}\right>_i \,,
\end{equation}
where $\left<...\right>$ represents a volume average. We present a derivation and discretized functional form for this expression in Appendix \ref{sec:irr_appendix}.

The spatial variation inherent to the radiative-heating term would create a temperature gradient in the dust and gas, and, via the dust's thermal emission, in $E_r$ as well. The resulting gas-pressure gradient would drive motions in the gas; moreover, the $\vec{F}_r$ driven by the radiation-energy gradient would couple to the dust momentum, and thus, to that of the gas as well (see e.g., Equation \ref{eq:momentum_exchange}). Over the long term, these two effects would disrupt the uniform-density initial condition and complicate interpretation of our results. In order to focus on thermal coupling, we therefore do not integrate the hydrodynamics block for this test. To allow radiation to leave the simulation box and prevent indefinite accumulation of radiation energy, we impose a fixed boundary at the left ($\vec{F}_r = 0$) and an outflow boundary ($\partial_x \vec{F}_r = 0$) at the right for radiative flux.

\begin{figure}
    \centering
    \includegraphics[width=1.0\linewidth]{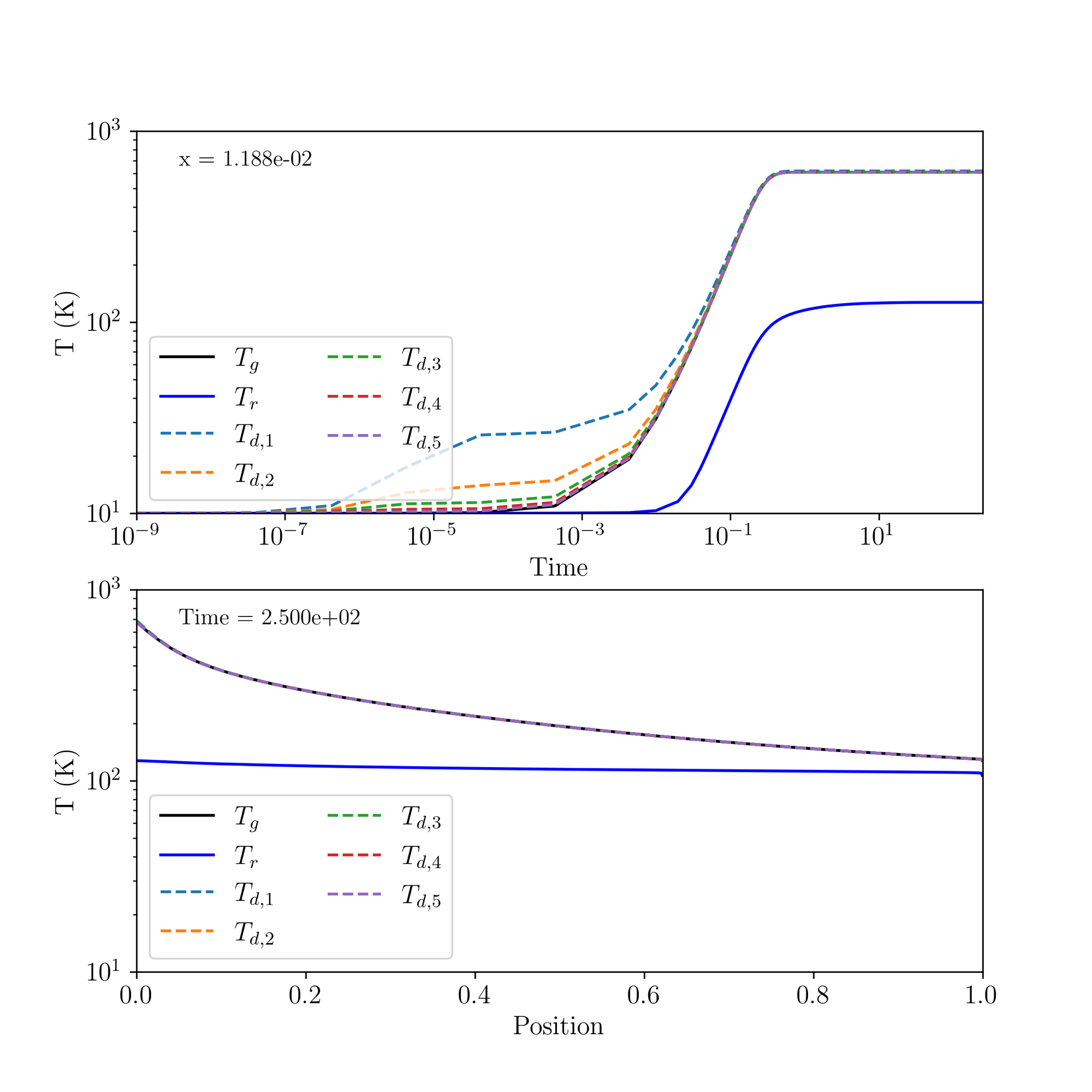}
    \caption{In the upper panel, we plot evolution at an x-position $x = 0.0118$, near the left of the domain. At early times, dust is rapidly heated by the irradiation flux, but then collisionally transfers its energy to the gas, and eventually, via emission, to the radiation field. In the lower panel, the gradient of dust/gas temperature leads to a gradient in the radiation temperature/energy density. This, in turn, drives a radiative flux which keeps the system in thermal balance even as the matter and radiation temperatures differ.}
    \label{fig:atmosphere_temperature_evolution}
\end{figure}

We run our simulation to a time $t_{\rm end} = 250 t_0$, by which point the temperatures of gas, dust, and radiation have stabilized everywhere. In Figure \ref{fig:atmosphere_temperature_evolution}, we plot the evolution of dust temperature over time at a position $x = 0.0118$, near the left-hand side of the domain. The dust species are the first to be heated, due to their high absorption opacities for ray-traced irradiation $(\kappa_{\rm irr})_{j,b}$, but gas-grain collisions rapidly cause this heating to be shared with the gas component. Over a longer timescale, the dust thermal-infrared opacities $\kappa_{d,j}$ become relevant, and cause the dust to emit into the radiation field. Because the irradiation term is ray-traced, the heating rate of the dust---and thus, the final dust temperatures---have a spatial gradient. The thermal emission rate, and thus the final radiation energy density $E_r$, therefore also show a spatial gradient. This, in turn, drives a radiative flux $\vec{F}_r$, whose divergence $\nabla \cdot \vec{F}_r$ contributes to the radiative energy balance (see Equation \ref{eq:er_evo}) and allows it to reach a steady state despite the inequality in the dust, gas, and radiation temperatures \citep[a common finding in systems with irradiation, e.g.,][]{MelonFuksman2021,Deng2023}. 

\begin{figure}
    \centering
    \includegraphics[width=1.0\linewidth]{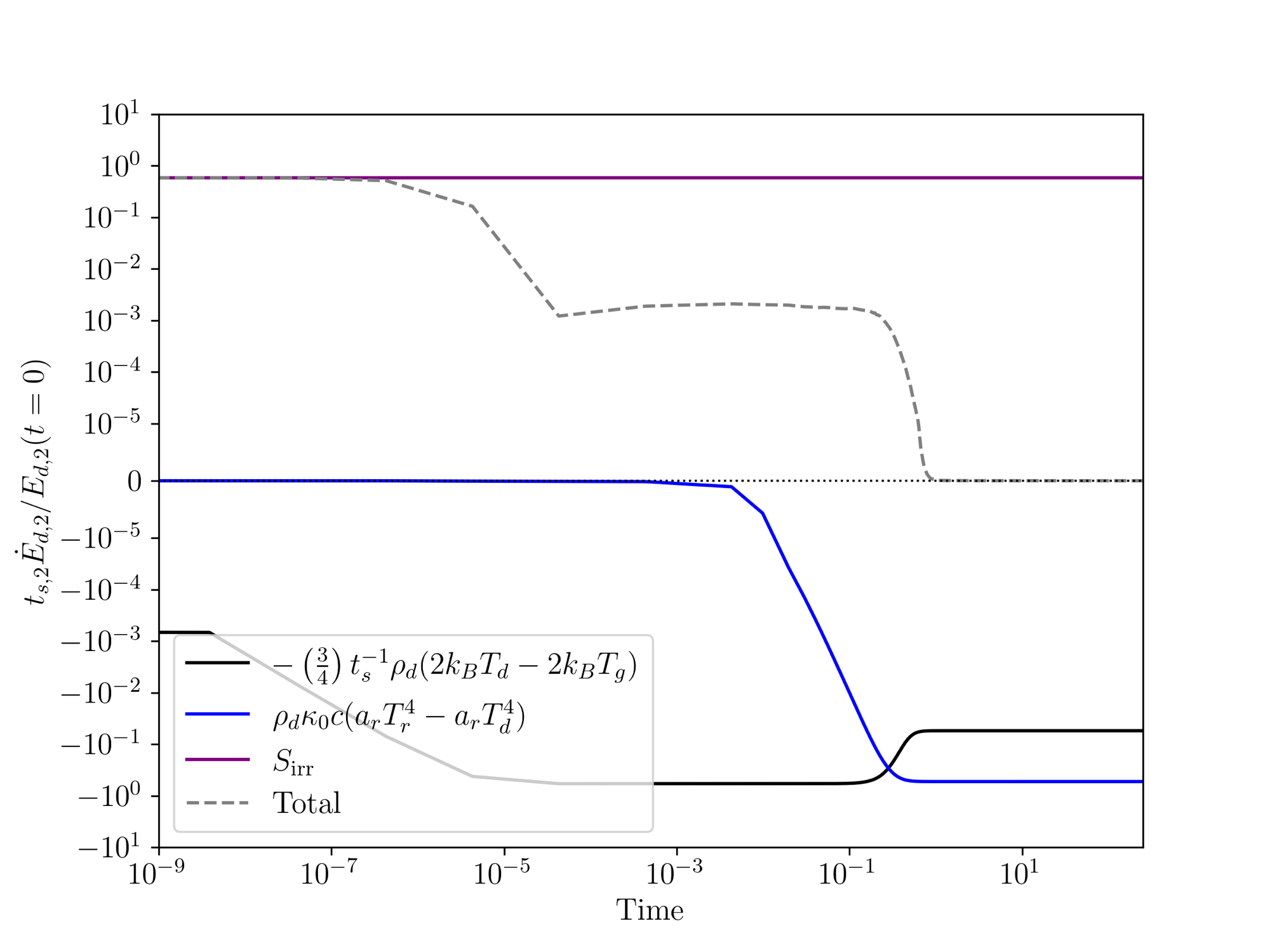}
    \caption{Energy balance for the dust species $j = 2$ at $x = 0.0118$. At late times, the sum of all terms equals zero and the system reaches steady state. The fact that the irradiation term $S_{\rm irr}$ is nonzero implies that, in equilibrium, the gas-dust coupling and radiation absorption/emission terms must be nonzero as well. This sustains temperature differences between gas, dust, and radiation even in steady state.}
    \label{fig:energy_balance_dust_2}
\end{figure}

To provide an alternative view of this process, and an additional check on our method, we plot the term-by-term energy balance for the dust species $j = 2$ in Figure \ref{fig:energy_balance_dust_2}. At the earliest times, the irradiation term dominates the dust's energy evolution. The resulting temperature difference between gas and dust drives the collisional term, which slows dust heating by several orders of magnitude. Regardless, the dust temperature eventually becomes sufficient to drive substantial thermal emission terms. Late in the simulation, the collisional and thermal-radiation terms settle on nonzero values, implying an enduring temperature difference between gas, dust, and radiation. However, the continued injection of ray-traced irradiation means that the sum of all three source terms nets to zero, demonstrating that the simulation accurately reproduces the true steady-state solution even when local thermodynamic equilibrium (LTE) between gas, dust, and radiation cannot be assumed.


\subsection{Radiative force on dust}
Radiation not only causes dust and gas to heat and cool, but also exerts forces on them, due to the absorption and scattering of the momentum contained within their photons. Among other astrophysical contexts, the radiation force on dust is particularly relevant to the irradiation instability at the inner edges of protoplanetary disks \citep{Fung2014,Bi2022}, as well as to the winds driven by asymptotic giant branch (AGB) stars \citep{Hoeffner2018}. 

To verify that our method can accurately handle the radiative force, we take a problem setup inspired by that of (the first problem). We modify the setup by setting all $t_{d,j} = 10^{10} t_0$, decoupling the dynamics of dust and gas, and setting all initial velocities to zero. We eliminate thermal coupling between dust and radiation by setting absorption opacities to zero, and set the scattering opacities $\sigma_{d,j} = \kappa_{d,0} \times 10^{-1} \times 0.5^{(j-1)}$. Furthermore, at $t = 0$, we set the radiative flux $F_{r} = +E_r$, implying that all of the radiation energy is initially contained within a rightward-traveling beam.

\begin{figure}
    \centering
    \includegraphics[width=1.0\linewidth]{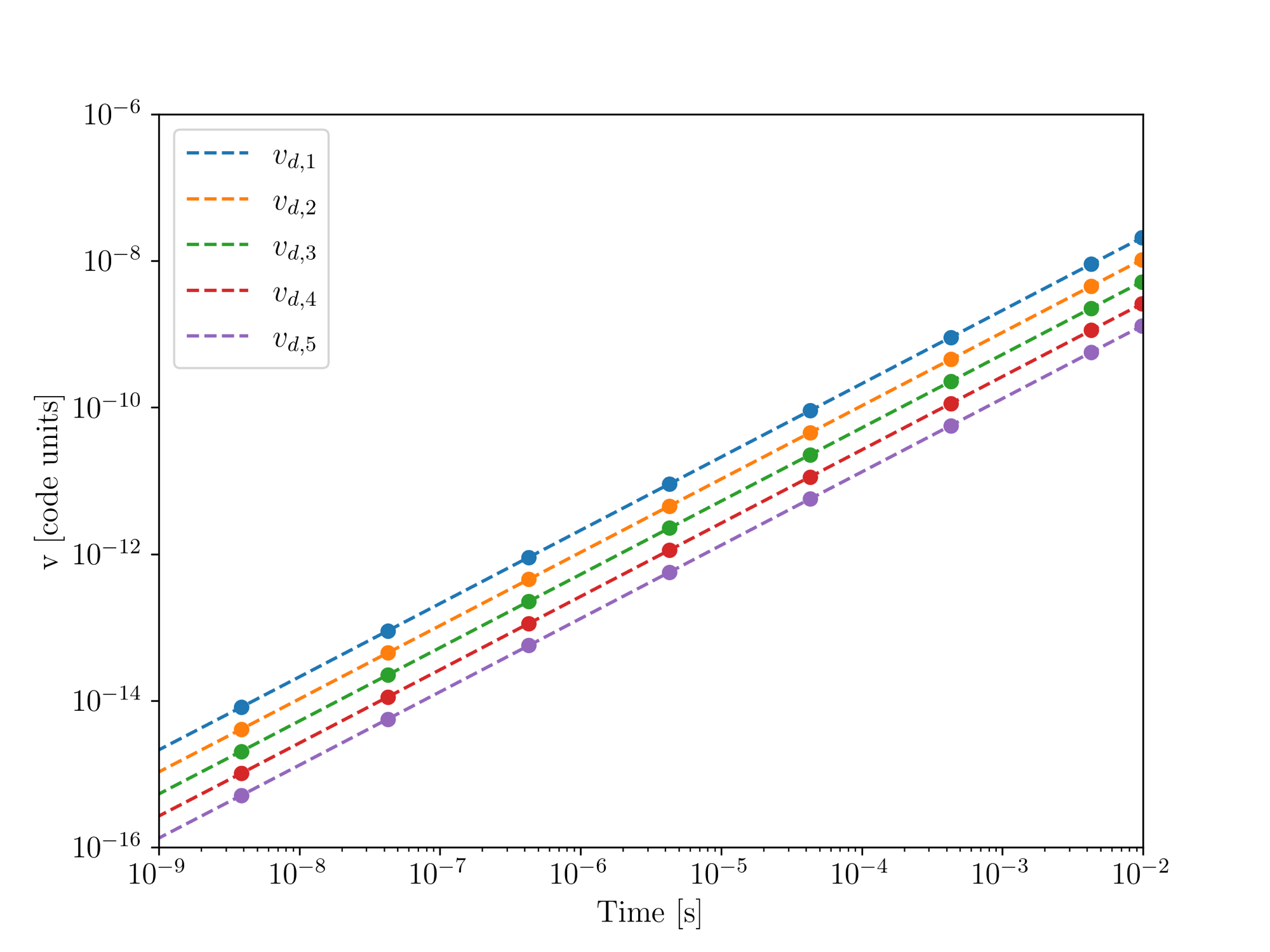}
    \caption{Evolution of the velocity for all dust species $v_{d,j}$ in the presence of a background radiative flux. Analytical solutions are in dotted line, while numerical solutions are plotted as filled circles.}
    \label{fig:dust_vx1_evolution}
\end{figure}

At early times, before the flux has been appreciably damped, the velocity of each grain species would be expected to evolve as $v_{d,j}(t) =F_r \sigma_{d,j} t$. In Figure \ref{fig:dust_vx1_evolution}, we plot the numerical evolution of dust velocity against this prediction, finding the relative deviation between them to be less than $10^{-3}$ at all times. 

\subsection{Multispecies dusty shock}\label{sec:dustyshock}
Shocks are ubiquitous in the context of protoplanetary disks, occurring as a consequence of planet-driven spiral density waves \citep{Goodman2001} and accretion of disk material onto planetary envelopes \citep[e.g.,][]{Tanigawa2012}, among other causes. The dynamics of dust in and around these shocks \citep{Sturm2020, Binkert2021, Krapp2024b}, in turn, exerts a substantial influence on disk observational appearance \citep{Speedie2022}. It is therefore important to verify that our method is well-behaved in the presence of shocks. \cite{Mattsson2016} and \cite{Lehmann2018} derived solutions for an isothermal shock including gas and a single dust fluid, the latter of which \cite{BenitezLlambay2019} subsequently generalized to incorporate multiple dust species \citep[see also e.g.,][]{Sewanou2025}:

\begin{subequations}
    \begin{equation}\label{eq:density_profile_gas}
        \rho_{g,0} \tilde{v}_{0} = \rho_{g} \tilde{v}_{g}
    \end{equation}
    \begin{equation}\label{eq:density_profile_dust}
        \rho_{d,j,0} \tilde{v}_{0} = \rho_{d,j} \tilde{v}_{d,j}
    \end{equation}
    \begin{equation}\label{eq:velocity_profile_gas}
        0 = \tilde{v}_g^2 + \tilde{v}_g\left(\sum_{j =1}^{n_d} f_{d,j,0}\left[\tilde{v}_{d,j} - \tilde{v}_{0}\right] - \left[\tilde{v}_{0}^{-1} + \tilde{v}_{0}\right]\right) + 1 \,
    \end{equation}    \begin{equation}\label{eq:velocity_profile_dust}
        \rho_{d,j}\tilde{v}_{d,j} \partial_x \tilde{v}_{d,j} = -\rho_{d,j} t_{d,j}^{-1}(\tilde{v}_{d,j} - \tilde{v}_{g})
    \end{equation}

The above relations are derived from the gas continuity, dust continuity, gas momentum, and dust momentum equations respectively, under the assumption of a steady-state shock. Zero-subscripts imply pre-shock values, and tildes indicate a velocity normalized to the sound speed $c_s$.\footnote{The functional form in \cite{BenitezLlambay2019}, presented in section 3.3.1, normalizes to the pre-shock velocity $v_0$ instead of to $c_s$. $\tilde{v}_0$ is equivalent to the shock Mach number, denoted by $\mathcal{M}$ in their work.} If we assume, as in \cite{BenitezLlambay2019}, that the coupling coefficient $K_{d,j} \equiv \rho_{d,j} t_{d,j}^{-1}$ in \Cref{eq:velocity_profile_dust} is a constant, the dust velocity profiles $\tilde{v}_{d,j}$ can be computed using an integrating factor, given an assumed functional form for $\tilde{v}_g$. Substituting the resulting $\tilde{v}_{d,j}$ into \Cref{eq:velocity_profile_gas} yields a revised estimate for $\tilde{v}_g$, which again can be substituted into \Cref{eq:velocity_profile_dust}; this procedure can be iterated to convergence to provide semi-analytic shock profiles. 

In the asymptotic post-shock limit, collisional coupling again drives dust and gas velocities to equality, $\tilde{v}_{g,1} = \tilde{v}_{d,j,1} = \tilde{v}_1$. It can be shown that
\begin{equation}\label{eq:asymptotic_vel}
    \tilde{v}_1 = \tilde{v}_0^{-1}\left(1 + \sum_{j=1}^{n_d} f_{d,j,0}\right)^{-1}
\end{equation}
\begin{equation}\label{eq:asymptotic_dens}
    \frac{\rho_{g,1}}{\rho_{g,0}} = \frac{\rho_{d,j,1}}{\rho_{d,j,0}} = \frac{\tilde{v}_{0}}{\tilde{v}_{1}} = \tilde{v}_0^2 \left(1 + \sum_{j=1}^{n_d} f_{d,j,0}\right)
\end{equation}
\end{subequations}
The additional factor of $\left(1 + \sum_{j=1}^{n_d} f_{d,j,0}\right)$ in the asymptotic scalings corresponds to the fact that, on timescales longer than those of dust-gas coupling, the dust and gas effectively behave as a single fluid, with an effective sound speed weighed down by the inertia of the pressureless dust component \citep{Laibe2014}.

For this test, we assume a left-state gas density $\rho_{g,0} = 1$, $\tilde{v}_0 = 2$, and $f_{d,j} = 0.5^{j-1}$ for dust species from 1 to 5, with coupling coefficients $K_{d,j} = 0.5^{j-1}$. Moreover, we set the global sound speed $c_s = 1$, with an adiabatic index $\gamma = 1.00001$ to ensure an essentially constant temperature and sound speed across the shock front. With this information, \Cref{eq:asymptotic_vel,eq:asymptotic_dens} can be used to evaluate the asymptotic right state. We place the shock jump at $x = 4$; for values $x \leq 4$, initial values are assigned based on the left state, whereas for $x > 4$, we use the asymptotic right state. Our computational domain, covered by $n_x = 1000$ cells, extends from $0 \leq x \leq 40$, with boundary values fixed to their initial conditions. We use an HLLC Riemann solver for the gas and a LeVeque solver for the dust.

\begin{figure}
    \centering
    \includegraphics[width=1.0\linewidth]{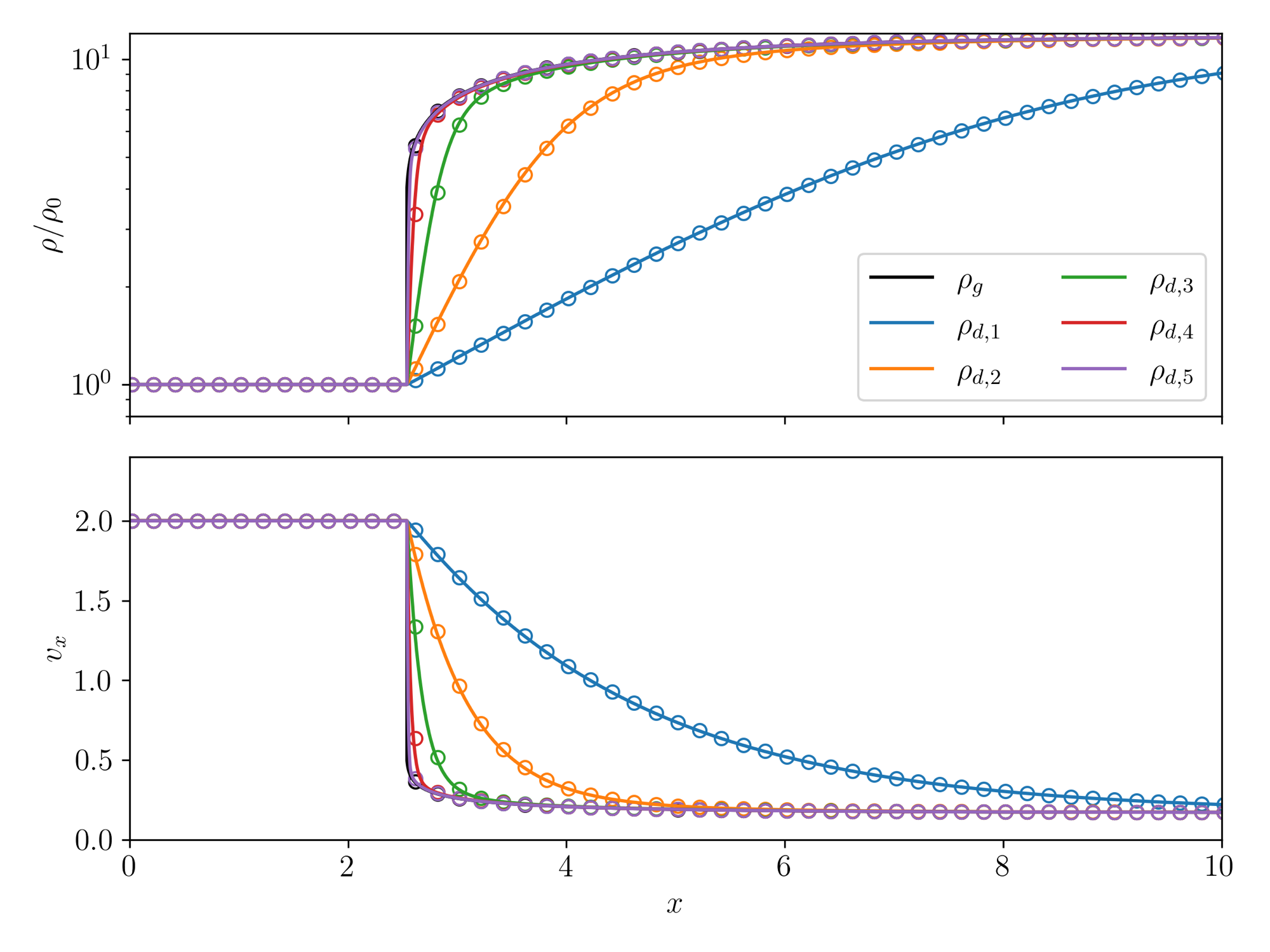}
    \caption{Results of the multispecies dusty-shock test described in \Cref{sec:dustyshock}; densities (rescaled to the initial density) are plotted in the upper panel, while velocities are plotted in the lower panel. We find close agreement between simulation outputs (open circles, sampled every 5 grid cells) and analytical theory (solid curves).}
    \label{fig:dustyshock}
\end{figure}

In \Cref{fig:dustyshock}, we present the results of our dusty-shock test at $t_{\rm end} = 500$, at which point the dust and gas profiles have achieved a steady state. Owing to its relatively low coupling coefficient $K_{d,1} = 1$ and high initial dust-to-gas ratio $f_{d,1,0} = 1$, dust species $j = 1$ experiences the slowest relaxation to the asymptotic density and velocity. By contrast, dust species $j = 5$, with a much stronger coupling $K_{d,1} = 2^4$ and lower dust-to-gas ratio $f_{d,5,0} = 2^{-4}$, very closely follows the trajectory of the gas. We find close agreement between simulations and analytical theory, with domain-averaged errors in the density and velocity of the various dust species ranging from 0.05-0.11\%.

%\dmnote{Run a test with an initial gas density 1, sound speed 1, and velocity 2, as is the case in Benitez-Llambay et al 2019. But try different dust fractions, can be very heavily loaded with 1, 1/2, 1/4, 1/8, 1/16. Can try coupling coefficients of 1, 2, 4, 8, 16.}

\begin{figure}
    \centering
    \includegraphics[width=1.0\linewidth]{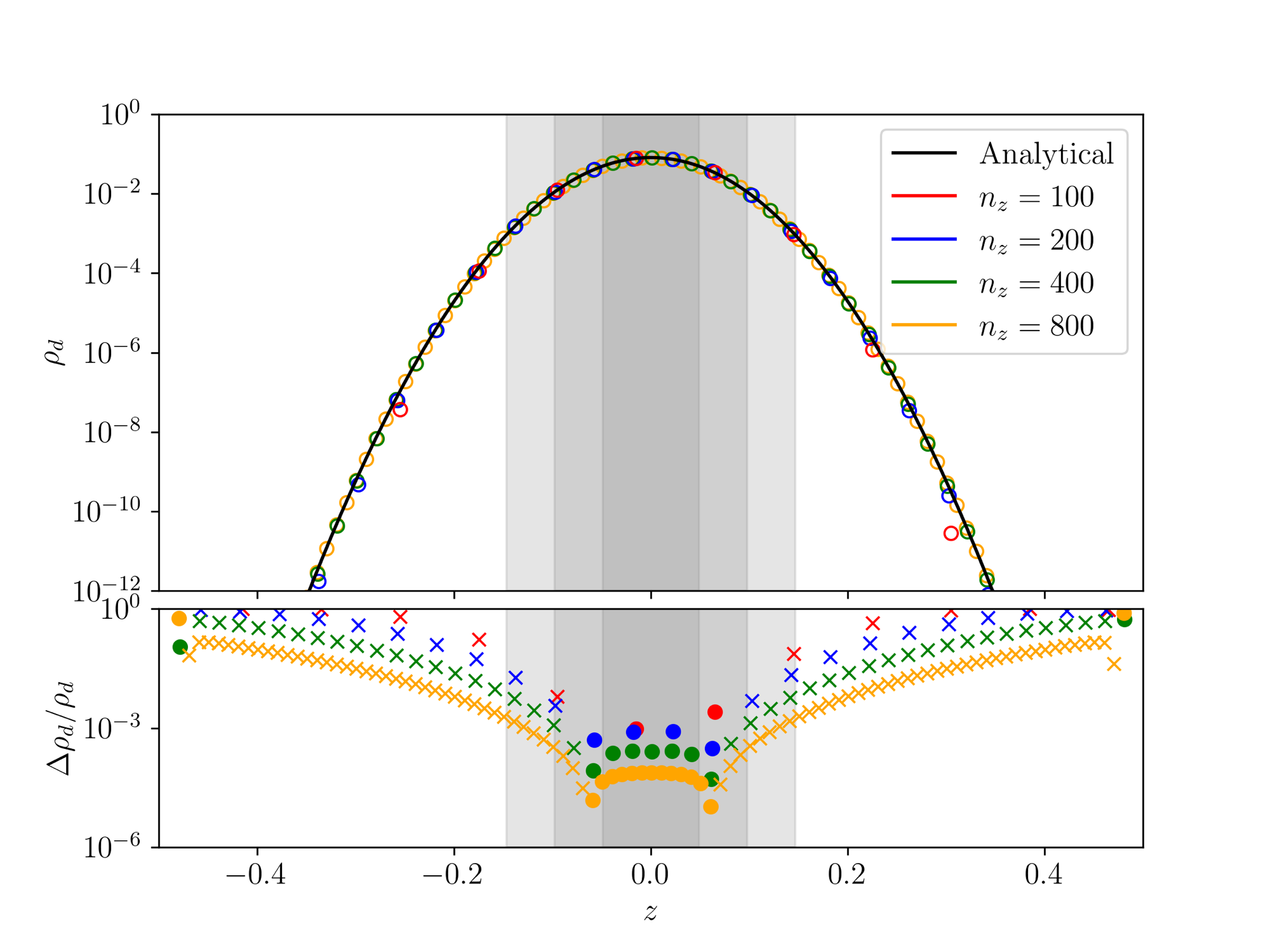}
    \caption{Graphical summary of results from the time-dependent diffusion test in \Cref{sec:diffusion_time_dep}. In the upper panel, the black curve indicates the analytical solution for dust density at $t_{\rm end} = 100$, whereas open circles indicate numerical values at various resolutions. In the lower panel, filled circles indicate positive relative errors, while crosses indicate negative errors. The shaded regions indicate the $\pm 1, \pm 2, \pm 3$ sd confidence intervals, containing 68\%, 95\%, and 99.7\% of the dust mass, respectively. Within these regions, errors are low, ensuring a standard deviation accurate to within ${\sim}0.5\%$ even at the coarsest resolution tested.}
    \label{fig:dynamic_diffusion}
\end{figure}

\subsection{Diffusion in a homogeneous medium}\label{sec:diffusion_time_dep}
Besides energy and momentum coupling, dust diffusion also plays a significant role in the structure and dynamics of protoplanetary disks. To verify the efficacy of our dust-diffusion module, we set up a test in which the gas has a constant uniform temperature and density, and in which the bulk velocities of both gas and dust are zero. Moreover, we assume no body force on either the gas or dust, and ignore the influence of radiation. Under these assumptions, all momentum equations (and the gas continuity equation) become identically zero on both sides, and the continuity equation for dust can be written in the form
\begin{equation}
    \frac{\partial \rho_{d,j}}{\partial t} + \nabla \cdot (\nu \nabla \rho_{d,j}) \,.
\end{equation}
We constrain the problem to one dimension, and take dust species $j$ to follow an initially Gaussian profile centered at $z = 0$, with  $\rho_{d,j,0}$ and dispersion $\sigma_{d,j,0}$ at $t = t_0$. Assuming that the diffusion coefficient is spatially and temporally constant, this setup admits the following analytical evolution for the time evolution of dust density:

\begin{equation}\label{eq:dust_diffusion_dynamic}
\begin{split}
    \rho_{d,j}(z, t) = &\rho_{d,j,0}\sqrt{\frac{\sigma_{d,j,0}^2}{\sigma_{d,j,0}^2 + 2\nu(t - t_0)}}\\
    &\exp\left(-\frac{z^2}{2(\sigma_{d,j,0}^2 + 2\nu(t - t_0))}\right) \,.
    \end{split}
\end{equation}
To study our scheme's handling of numerical diffusion, we set up a 1D grid over the domain $x \subseteq [-0.5,0.5]$. We assume a constant background gas density $\rho_g = 1$ and velocity $v_{g,z} = 0$, and fix $c_{s, \rm iso} = 0.1$ with an ideal-gas $\gamma = 1.4$. We assume that the single dust species we simulate follows an initial profile described by \Cref{eq:dust_diffusion_dynamic}, with $\rho_{d,1,0} = 0.01 \sqrt{1/2\pi\sigma_{d,1,0}^2}$, $\sigma_{d,1,0} = 0.02$, and $t_0 = 0$; the initial dust velocity is set to zero, and the stopping time $t_{s,1} = 10^{-10}$. To compute the diffusion coefficient $\nu$, we set the viscous $\alpha = 10^{-3}$; in the absence of a gravitational field and thus a dynamical time/orbital frequency, we assign the turbulent turnover time $\Omega_{\rm turb} = 1$. We test four different numerical resolutions, $n_z = \{100,200,400,800\}$ grid cells, picking $n_z = 200$ as the fiducial case.

\Cref{fig:dynamic_diffusion} presents the results of our dynamic-diffusion test at $t_{\rm end} = 100$, by which point the initial Gaussian dust packet has broadened to $\sigma_{d,1}(t_{\rm end}) = 0.04898$. We find that the numerical and analytical solutions agree well with one another across many orders of magnitude in density. Relative density errors tend to be small and positive near the Gaussian center, but larger and negative in the wings. We attribute this behavior to truncation error in the face-centered evaluation of diffusion velocity (\Cref{sec:advection_diffusion}), which is larger in regions (such as the Gaussian wings) where $\nabla \ln (\rho_d/\rho_g)$ is steeper. Nevertheless, the low densities in these regions mean that such errors have little influence on the dust packet in aggregate. At the resolutions tested, the numerical and analytical standard deviations differ only by $\{-0.479\%,-0.184\%,-0.059\%,-0.015\%\}$, respectively.

\subsection{Vertical dust settling}\label{sec:settling_diffusion}
In a protoplanetary disk, the vertical structure of gas is determined by the balance between the gravity, which draws material toward the midplane, and pressure support, which counteracts this force. Assuming a pressure profile $p_g = \rho_g k_B T/\mu = \rho_g c_{s, \rm iso}^2$ where $T$ is independent of the vertical position $z$, and in the limit that $z \ll R$, it can be shown that the vertical gravitational acceleration is given by
\begin{equation}
    (\nabla \Phi) \cdot \hat{z} = -GM_* (R^2 + z^2)^{-3/2} z \approx -\Omega_K^2 z \,.
\end{equation}
where $M_*$ is the stellar mass, $\Omega_K$ the local Keplerian orbital frequency, and $c_{s, \rm iso} = c_s/\gamma$ an ``isothermal sound speed'' which incorporates the conversion between pressure and density. Setting the gravitational acceleration equal and opposite to the pressure gradient $\nabla p_g = c_{s, \rm iso}\nabla \rho_g$, and integrating, yields the classical Gaussian vertical profile of protoplanetary disks \citep[e.g.,][]{Armitage2020book}:
\begin{equation}\label{eq:gas_gaussian}
    \rho_g = \frac{\Sigma_{g,0}}{\sqrt{2\pi H^2}} \exp\left(-z^2/2H^2\right)
\end{equation}
where in this case, the scale height $H = c_{s, \rm iso}/\Omega_K$.

Unlike the gas, dust does not receive pressure support, but does experience drag with the gas. Balancing these two forces shows that the dust will fall to the midplane at a terminal velocity
\begin{equation}\label{eq:terminal_velocity_dust}
    -\rho_{d.j}\Omega_K^2 z + \rho_{d.j}t_{d,j}^{-1}(v_{d,j,z} - v_{g,z}) = 0 \rightarrow v_{d,j,z} = \Omega_K z \St_j
\end{equation}
where we set $v_{g,z} = 0$ and define the Stokes number as the ratio of the stopping time and the dynamical time, $\St_j \equiv t_{d,j} \Omega_K$. In steady state, this velocity is counteracted by an equal and opposite diffusion velocity. Setting the total vertical dust velocity $v_{d,j,z} + v_{d,{\rm diff},j,z} = 0$, and substituting in the functional forms, yields the following:
\begin{equation}\label{eq:dust_concentration}
\begin{split}
    \Omega_K z\St_j
    - \nu \nabla \ln\left(\frac{\rho_{d,j}}{\rho_g}\right)
    ={}& 0 \\
    \;\longrightarrow\;
    \frac{\rho_{d,j}}{\rho_g}
    ={}&
    \left(\frac{\rho_{d,j}}{\rho_g}\right)_0
    %&\times
    \exp\left(
        -\frac{\Omega_K z^2\St_j}{2\nu}
    \right)
    \\
    ={}&
    \epsilon_{d,j}
    \sqrt{1+\frac{\St_j}{\alpha}}
    \exp\left(
        -\frac{z^2}
        {2(\alpha/\St_j)H^2}
    \right)\,,
\end{split}
\end{equation}

where, as is common in protoplanetary disk studies, we have assumed that $D = \nu = \alpha c_{s, \rm iso} H_g$ given that the dust scale height $H_g \equiv c_{s, \rm iso} \Omega_K^{-1}$. $(\epsilon_{d,j})$ represents the column-integrated dust-to-gas ratio. From Equations \ref{eq:gas_gaussian} and \ref{eq:dust_concentration}, we find that the dust scale height is
\begin{equation}
    H_{d,j} = \frac{H_g}{\sqrt{1+\St_j/\alpha}}\,.
\end{equation}

When $\St_j/\alpha \gg 1$, dust tends to form a thin layer at the disk midplane; when $\St_j/\alpha \ll 1$, it is evenly dispersed throughout the disk gas. When multiple grain sizes are present, this causes the grain size distribution to vary as a function of vertical position, and with it, properties such as the (frequency- and distribution-integrated) opacities and gas thermal relaxation times \citep[e.g.,][]{Muley2023,Pfeil2024}.

To study our scheme's ability to capture vertical diffusion, we set up a 1D grid over the domain $z \subseteq [-0.5,0.5]$. We assume a gravitational field $(\nabla \Phi) \cdot \hat{z} = -\Omega_K^2 z$, using a system of units in which $\Omega_K = \Sigma_{g,0} = 1$. Rather than evolving the radiation consistently, we fix $c_{s, \rm iso} = 0.1$ with an ideal-gas $\gamma = 1.4$. We set the initial $v_{g,z} = 0$ everywhere in the domain, with a viscous $\alpha = 10^{-3}$.

Assuming hydrostatic equilibrium yields a Gaussian vertical density profile (Equation \ref{eq:gas_gaussian}) with $H = 0.1$, which we use as the initial condition for $\rho_g$. The densities of the $n_d = 7$ dust species are initialized to the same profile as the gas, scaled by a global factor $f_{\rm dg, j} = 10^{-2} \times 0.5^{(j-1)}$ and with $\St_j \equiv 0.25^{(j-1)}$. We test the same four different numerical resolutions, $n_z = \{100,200,400,800\}$, as in \Cref{sec:diffusion_time_dep}, corresponding to 10, 20, 40, and 80 cells per gas scale height (cps) respectively; we choose $n_z = 200$ as our fiducial resolution. Throughout the duration of the simulation, we fix gas pressure, density, and velocity equal to their initial values. We run all simulations to $t_{\rm end} = 1000 \Omega_K^{-1}$, corresponding to ${\sim}159$ orbits of the simulated disk column.

\begin{figure}
    \centering
    \includegraphics[width=1.0\linewidth]{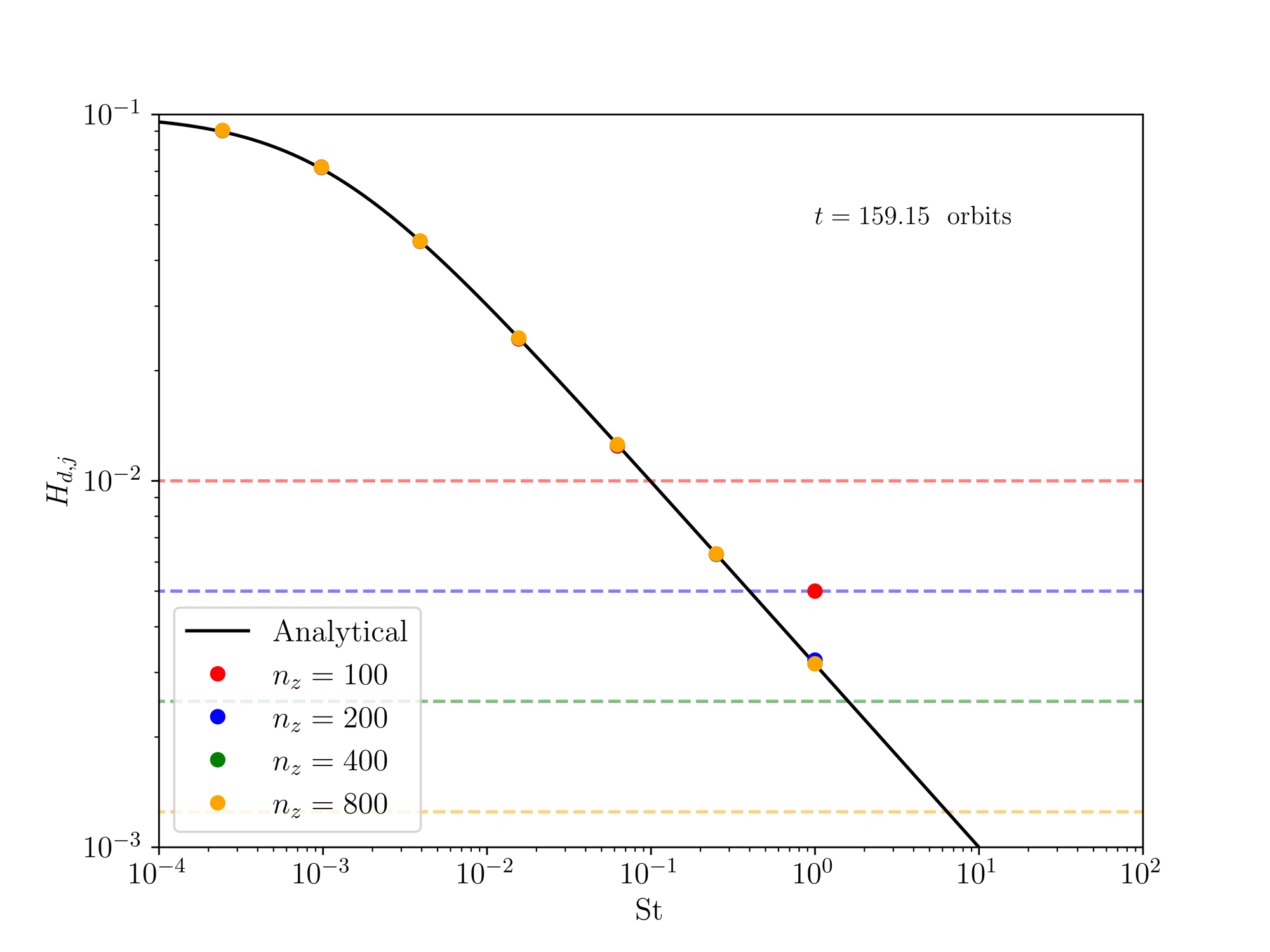}
    \caption{Dust vertical scale height  $H_{d,j}$ as a function of $\St$ (for $\alpha = 0.001$ and $H_g = 0.1$) at various numerical resolutions, along with the analytical solution (black) and the grid scale at each resolution (dotted lines). Agreement between numerics and analytics is excellent across all resolutions and Stokes numbers, save for $n_z = 100$ at $\St = 1$, where the thickness of the dust layer is severely underresolved.
    }
    \label{fig:vertical_diffusion}
\end{figure}

Over the course of the simulation, settling and diffusion act to drive the vertical profile of each dust species toward a steady state. At the end, we compute the scale height for each species as
\begin{equation}
    H_{d,j}^2 = \left<(z - \left<z\right>)^2\right> \equiv \frac{1}{\Sigma_{d,j}}\sum_{i}(z_i - \left<z\right>)^2 \rho_{d,j} \Delta z_i \,.
\end{equation}
where $\left<z\right> \equiv (\Sigma_{d,j}^{-1}) \sum_{i}(z_i - \left<z\right>)^2 \rho_{d,j} \Delta z_i$, consistent with zero for our setup. In Figure \ref{fig:vertical_diffusion}, we plot these simulation-derived dust scale heights as a function of Stokes number for each of the resolutions we test. These values are compared to the analytical prediction, shown in black, and to the size of a single grid cell, shown in dotted lines color-coded by resolution. For all Stokes numbers and resolutions, we find very close agreement between the simulated and analytical values---the sole exception being for $n_z = 100$, where the analytical scale height for $\St = 1$ is overestimated by a factor of ${\sim}1.5$.

Although our diffusion scheme itself is stable and accurate, we do find some deviation between simulations and analytics when the gas is allowed to evolve self-consistently. In these simulations (not shown), spatial discretization errors in the calculation of hydrostatic equilibrium drive spurious velocities, particularly at higher altitudes. For fine grids and high-$\St$ grains, in which spatial errors are small and dust terminal velocities are large, these effects are unimportant. For coarse grids and low $\St$---in other words, when spatial error is comparatively high and terminal settling velocity relatively low---the impact of these spurious velocities on the settling-diffusion equilibrium, and thus the dust layer thickness, can be substantial. When running dynamical simulations, we advise the reader to minimize these spurious velocities by ensuring that the initial conditions are numerically in hydrostatic balance, and by using damping layers at the upper and lower boundaries.

\begin{figure}
    \centering
    \includegraphics[width=1.0\linewidth]{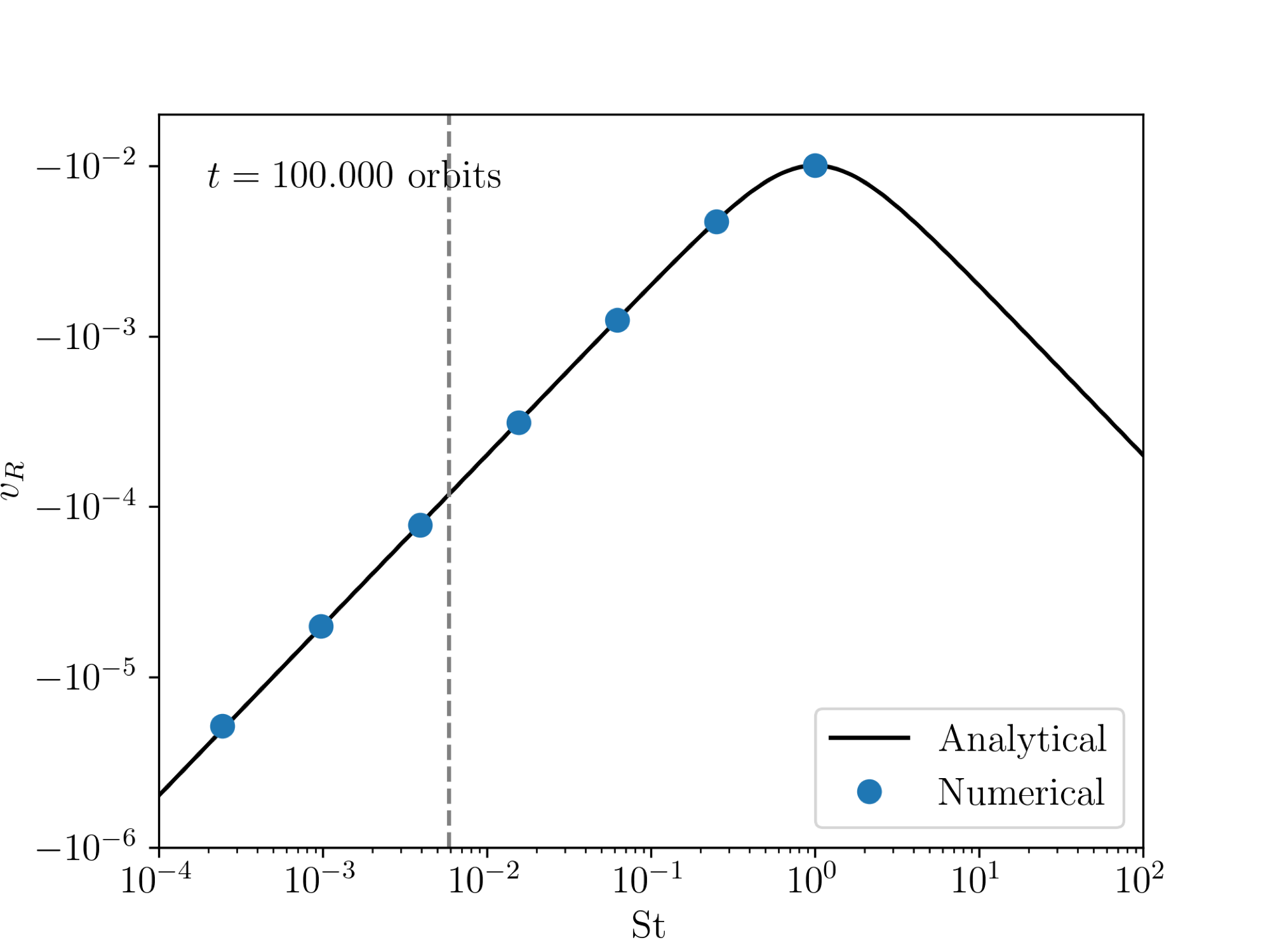}
    \caption{Radial drift velocity as a function of $\St$ at $R = 1$, with numerically obtained points for all dust species in blue plotted against the ($v_{d,r} \approx 2(v_{\phi,g} - v_K)/(\St + \St^{-1})$) in black solid line. The vertical dotted line indicates the dimensionless numerical timestep $\delta t \Omega_K$. The largest error in $v_{d, R}$ is 5.61\% for $j = 7$ ($\St = 0.25^{(7-1)}$); for all other species, the error is 1.01\% or less.}
    \label{fig:radial_drift_comparison}
\end{figure}

\begin{figure}
    \centering
    \includegraphics[width=1.0\linewidth]{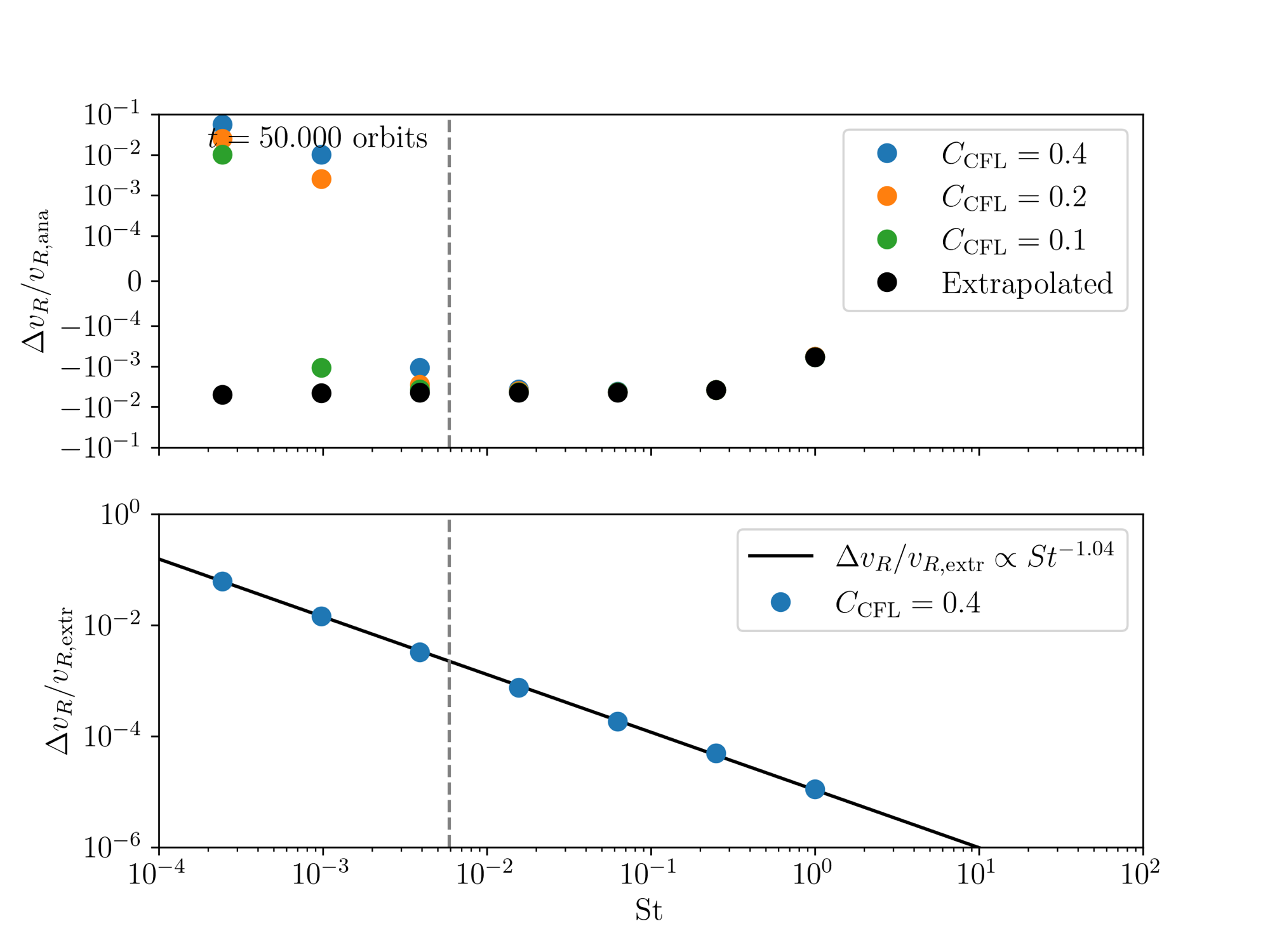}
    \caption{Error properties of the radial drift simulation. In the upper panel, we show the deviation of simulated radial drift velocities from the analytical estimate as a function of Courant number, as well as an extrapolated estimate for the true solution (see text for details). In the lower panel, we show the estimated deviation between the numerical radial drift velocities for all dust species (using the fiducial $C_{\rm CFL} = 0.4$) and the extrapolated estimate of the true solution; the best-fit slope for these points is consistent with first-order convergence in $\Delta t \Omega_K/\St$.}
    \label{fig:radial_drift_error}
\end{figure}

\subsection{Radial drift}\label{sec:radial_drift}

Gas in a protoplanetary disk experiences pressure forces, causing its orbital velocity to deviate from that of the dust, which does not. The gas-dust velocity difference---negative if the radial pressure gradient is negative, and vice versa---creates a mutual drag force in the azimuthal direction, affects the dust's angular momentum and leads to a radial dust drift with a speed (see Appendix \ref{sec:ring_trapping} for more details)
\begin{equation}\label{eq:radial_drift}
    v_{d,r} \approx \frac{2(v_{g, \phi} - v_K)}{\St + \St^{-1}} \approx \frac{{v}_{\rm K} h^2}{\St + \St^{-1}} \frac{\partial \ln P}{\partial \ln R} \,,
\end{equation}
where $h$ is the local scale height ratio, $v_K$ the Keplerian orbital velocity, $P$ the pressure, $R$ the cylindrical radius. For large bodies ($\rm St \gg 1$), the dust-gas drag force is weak enough that angular momentum loss, and thus radial drift, is very slow. For the smallest grains, coupling is so strong that the dust becomes entrained in the gas, damping the rate of radial drift to zero.\footnote{Strictly speaking, in the limit of small $\St$, all components of the dust and gas velocity would be equal, i.e., any radial motion in the gas due to viscosity or disk-planet interaction would be immediately followed by the dust.} It is for intermediate-sized solids ($\St \approx 1$) that radial drift proceeds with maximal efficiency. 

For properties typical of the solar nebula at one Earth radius $R = 1 {\rm \ au}$, an object of $\St \approx 1$ would have a physical radius of $a_{\rm gr} \approx 1 {\rm \ m}$ \citep{Armitage2020book} The radial-drift timescale $t_d \equiv R/v_{d,r} \approx 10^2 {\rm \ yr}$ is much shorter than the grain-growth timescale, implying that meter-sized boulders would be swept toward the central star long before growing into planetesimals (in the high-\St \ regime). How this ``meter-sized barrier'' may be overcome---for instance, by trapping such $\St \approx 1$ particles in substructures such as rings and vortices---is a major open question in protoplanetary disk studies, making a radial-drift test problem especially relevant for our work. From the numerical point of view, such a test would verify the correct implementation of the geometric source terms in curvilinear coordinates.

To study radial drift, we initialize a 2D cylindrical grid in \texttt{PLUTO} over the domain $R \subseteq [0.4, 2.5]$ and $\phi \subseteq [0, 2\pi)$. We use a resolution of 256 ($R$) $\times$ 877 ($\phi$) cells, spaced logarithmically in $R$ and linearly in $\hat{\phi}$. We use an initial surface density of 
\begin{equation}\label{eq:sigma_bg}
    \Sigma_g(t = 0) = \Sigma_0 (R/R_0)^{-1} \,,
\end{equation}
where $\Sigma_0 = R_0 = G = M_* = \Omega_K^{-1} = 1$ in code units. We use a constant disk aspect ratio $h = H/R = 0.1$, where $H$ is the absolute disk scale height. We use $n_d = 7$ dust species, all initialized with a density $\Sigma_{d,j}(t = 0) = f_{\rm dg} \Sigma_g(t = 0)$, corresponding a dust-to-gas ratio $f_{\rm dg} = 10^{-4}$. Stopping times are defined such that each dust species has a constant Stokes number throughout the disk, $\St_j \equiv 1 \times 0.25^j$; given that the fluid approximation is expected to break down for $\St \gtrsim 1$, we do not test this regime.

We impose the following initial velocity field on the gas and all dust species:
\begin{equation}
\begin{split}
    v_r &= 0 \\
    v_{\phi} &= \sqrt{v_K^2 + c_s^2 \frac{\partial \ln P}{\partial \ln R}}
\end{split}
\end{equation}
where $v_{\phi}$ accounts for the pressure support the gas receives. To ensure the stability of our integration, we use a low \citep{Shakura1976} $\alpha = 10^{-6}$, where $\nu \equiv \alpha c_s H$, for both dust and gas diffusion.

We employ periodic boundary conditions in the azimuthal direction. At the radial boundaries, we fix all fluid fields to their initial condition, with the exception of the dust radial drift velocities, for which we use the analytical value given by Equation \ref{eq:radial_drift}. To further mitigate the impact of wave reflection on our results, we reset all gas variables to their initial values at each timestep. We use a fiducial CFL factor $C_{\rm CFL} = 0.4$. We accelerate our simulation using \texttt{PLUTO}'s implementation \citep{Mignone2012} of the \texttt{FARGO} orbital advection scheme \citep{Masset2000}, which rotates the grid at Keplerian velocity and lifts the stringent limitation that it would otherwise impose on the numerical timestep.

In Figure \ref{fig:radial_drift_comparison}, we plot the dust radial-drift velocities computed from our simulations against the analytical prediction from Equation \ref{eq:radial_drift}. Whether or not the timestep resolves the dust stopping time ($\St > \Delta t \Omega_K$), we find that our scheme is numerically stable and produces results that agree with the analytical expectation; this is an intended and desirable property of our implicit-explicit scheme. Nevertheless, there does remain a small amount of error, growing to $\sim 5.6\%$ for the smallest $\St$ that we test.

To more rigorously understand the error properties of our numerical scheme, we must disentangle the error produced by the scheme itself from that introduced by the analytical solution, which is itself only accurate to first order in the relative gas-velocity deviation,  $(v_{g, \phi} - v_K)/v_K \sim h^2$. To do so, we run two additional simulations at $C_{\rm CFL} = 0.2$ and $C_{\rm CFL} = 0.1$, corresponding to a numerical timestep one-half or one-quarter the fiducial value, respectively. We then compute the ratio 
\begin{equation}
    s_{\rm conv} = \frac{\left[v_{d,j,R}(C_{\rm CFL} = 0.2) - v_{d,j,R}(C_{\rm CFL} = 0.1)\right]}{\left[v_{d,j,R}(C_{\rm CFL} = 0.4) - v_{d,j,R}(C_{\rm CFL} = 0.2)\right]} \,,
\end{equation}
finding that $s_{\rm conv} \approx 0.50$. We use this ratio to compute an infinite sum and extrapolate the ``true'' drift velocity in the limit that $C_{\rm CFL} \rightarrow 0$:
\begin{equation}
\begin{split}
    v_{d,j,R, \rm extr} = &-\left[v_{d,j,R}(C_{\rm CFL} = 0.4) - v_{d,j,R}(C_{\rm CFL} = 0.2)\right]\\
    &\sum_{k = 0}^{\infty} s_{\rm conv}^k + v_{d,j,R}(C_{\rm CFL} = 0.4)
    \end{split}
\end{equation}
where to evaluate the sum, we make use of the property $(1 - s_{\rm conv})^{-1} = \sum_{k = 0}^{\infty} s_{\rm conv}^k$. We show the results of this procedure in the upper panel of Figure \ref{fig:radial_drift_error}. In the limit of low Stokes number, the relative deviation between the analytical and extrapolated drift velocities $\Delta v_{R,  \rm extr}/v_{R, \rm ana} \approx -10^{-2} \approx -h^2 \approx (v_{g, \phi} - v_K)/v_K$, consistent with the notion that it is caused by $\mathcal{O}(((v_{g, \phi} - v_K)/v_K)^2)$ terms unaccounted for by Equation \ref{eq:radial_drift}. In the lower panel of Figure \ref{fig:radial_drift_error}, we plot the relative difference between the numerical (at the fiducial $C_{\rm CFL} = 0.4$) and extrapolated radial velocities; for tightly-coupled grains, errors are quite large, but decline linearly with increasing $\St$. This, combined with the fact that the scaling of error for the same $\St$ at different timesteps also appears to be linear ($s_{\rm conv} \approx 0.5 \approx (0.2/0.4)^1)$, demonstrates our code to be first-order accurate when implicit source terms are important. We note that this procedure only corrects for error introduced by the temporal discretization $\Delta t$; those caused purely by the spatial discretization $\Delta R, R\Delta \phi$ (e.g., in spatial reconstruction or force evaluation) remain untouched.

\subsection{Radial dust trapping}\label{sec:radial_trapping}
Equation \ref{eq:radial_drift} implies that radial drift should stop when pressure reaches a local maximum---for instance, at the outer edge of a planet-opened gap, or at a ``traffic jam'' created by the viscosity transition at the edge of the MRI-active zone. Over time, this would create bright, observable dust rings \citep[e.g.,][]{Dullemond2018} which may serve as sites for subsequent planet formation. As in the vertical direction, this radial accumulation would be balanced out by concentration diffusion. The final width of the dust ring can be estimated by setting $v_{d,j,R} + v_{d,{\rm diff},j,R} = 0$ at the location of the pressure peak $R = R_{\rm max}$ (following Appendix \ref{sec:ring_trapping}):
\begin{comment}
\begin{equation}
    w_{dg} \equiv \left[-\alpha ({\St_j + \St_j^{-1}})/(P^{-1} \partial^2_RP)_{R_{\rm max}}\right]^{1/2} = \left[-\alpha ({\St_j + \St_j^{-1}})/(R^{-1} \partial_R(\partial \ln P/\partial \ln R)_{R_{\rm max}}\right]^{1/2}
\end{equation}

from which the concentration of dust grains within the pressure bump can be evaluated as:
\begin{equation}
    \frac{\epsilon_{j}(R)}{\epsilon_j(R_{\rm max})} \approx \exp(-(R - R_{\rm max})^2/2w_{dg}^2)
\end{equation}
\end{comment}

For this test, we use the setup from Section \ref{sec:radial_drift} as a starting point, but change the density profile to be dominated by a pressure bump

\begin{equation}
    \Sigma_{g, \rm bump} = 2.5 \Sigma_0 \exp\left(-(R - R_{\rm max})^2/2w_{\rm ring}^2\right)
\end{equation}

with the total initial gas density $\Sigma_{g}(t = 0) = 10^{-5} \Sigma_{g, \rm bg} + \Sigma_{\rm bump}$, where $\Sigma_{g, \rm bg} = \Sigma_0 (R/R_0)^{-1}$ as in Equation \ref{eq:sigma_bg}. We set the sound speed to a constant $c_s = 0.1$ code units throughout the domain. For each of the $n_d = 7$ dust species, we set an initial $f_{\rm dg} = 10^{-9}$; we use the same Stokes numbers as in Sections \ref{sec:settling_diffusion} and \ref{sec:radial_drift}, but set $\St_7 = 10^{-10}$ to provide a basis for comparison. Owing to the long timescales required to achieve drift-diffusion equilibrium, we opt for an axisymmetric 1.5D ($R-\phi$) setup with 256 radial cells; throughout the integration, we freeze the gas background to its initial condition. 

\begin{figure}
    \centering
    \includegraphics[width=1.0\linewidth]{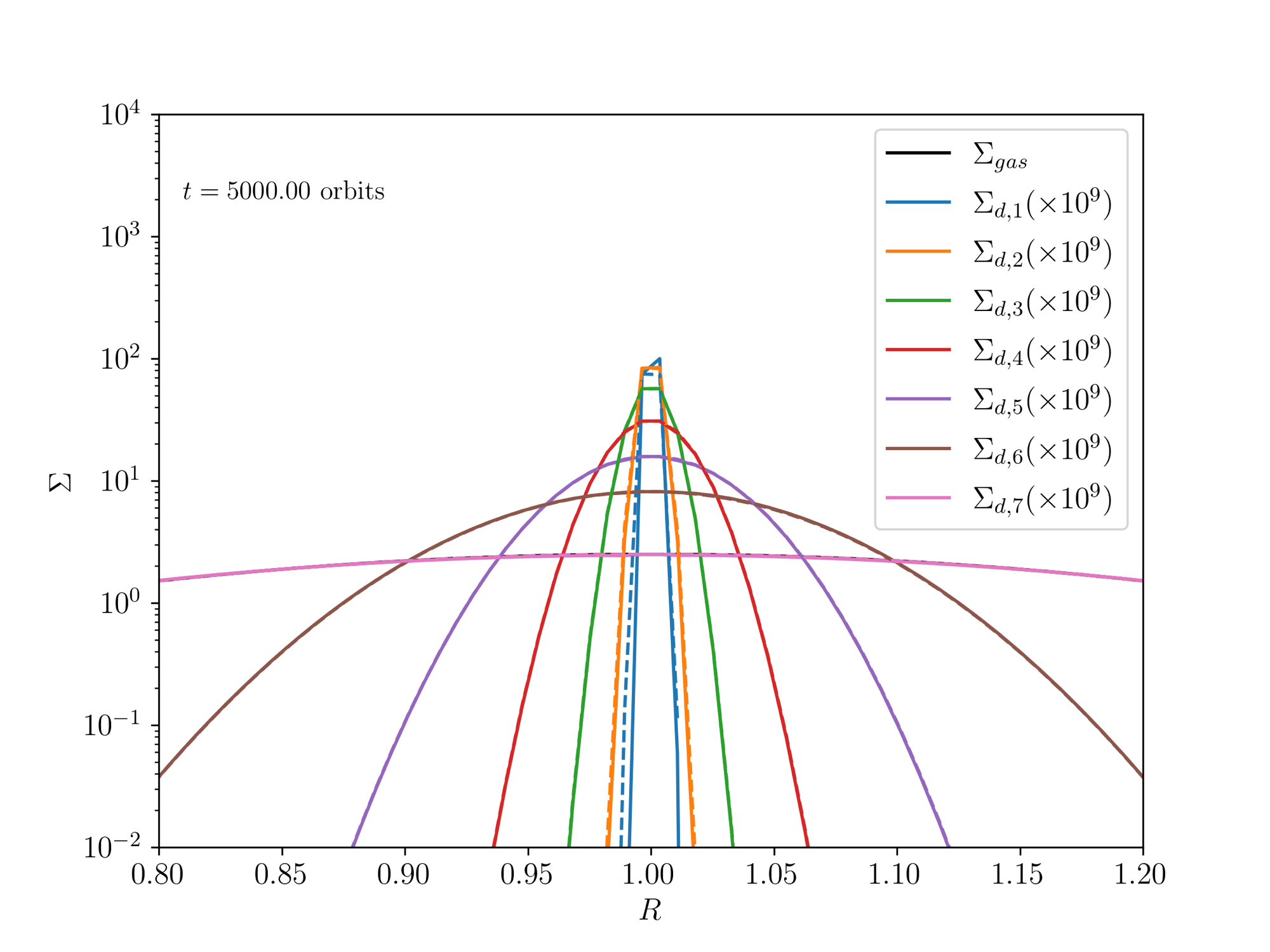}
    \caption{Gas (solid black) and dust (solid colors, scaled up by $10^6$ for visibility) densities at $t_{\rm end} = 3000 \times 2\pi \Omega_K^{-1}$ in the radial dust-trap test. We compare these simulated dust profiles to semi-analytical predictions (dashed lines), and find them to be virtually indistinguishable in most cases. For dust species $j = 1$ and $j = 2$, the ring width $w_{\rm dg}$ is resolved by less than one grid cell, leading to deviations between the predicted and actual ring profiles. Dust species $j = 7$, which is well-coupled with the gas, maintains an essentially constant dust-to-gas ratio throughout the simulation, leading it to cover up the black $\Sigma_{\rm gas}$ curve.}
    \label{fig:gaussian_bump}
\end{figure}

In Figure \ref{fig:gaussian_bump}, we plot (with solid lines) dust and gas densities in our simulation at $t = 5000$ orbits. Alongside these, we include (with dashed lines) semi-analytical predictions for the dust-ring profile, obtained by setting the radial-drift velocity in Equation \ref{eq:radial_drift} equal and opposite to the dust concentration-diffusion velocity from Equation \ref{eq:diffusion}, and then numerically integrating and normalizing (see also Appendix \ref{sec:ring_trapping}). We observe some deviation between the numerical and analytical profiles for $\Sigma_{d,1}$, and to a lesser extent for $\Sigma_{d,2}$, because the ring width $w_{\rm dg}$ in these cases is resolved by less than one grid cell $\Delta R(R = 1)$. Already for $\Sigma_{d,2}$, for which $w_{\rm dg} \approx 1.2 \Delta R(R = 1)$, we observe a near-perfect agreement between the numerical and analytical predictions, with the curves becoming visually indistinguishable from one another.

\section{Application: Dust drift and sublimation}\label{sec:sublimation}
\begin{figure}
    \centering
    \includegraphics[width=1.0\linewidth]{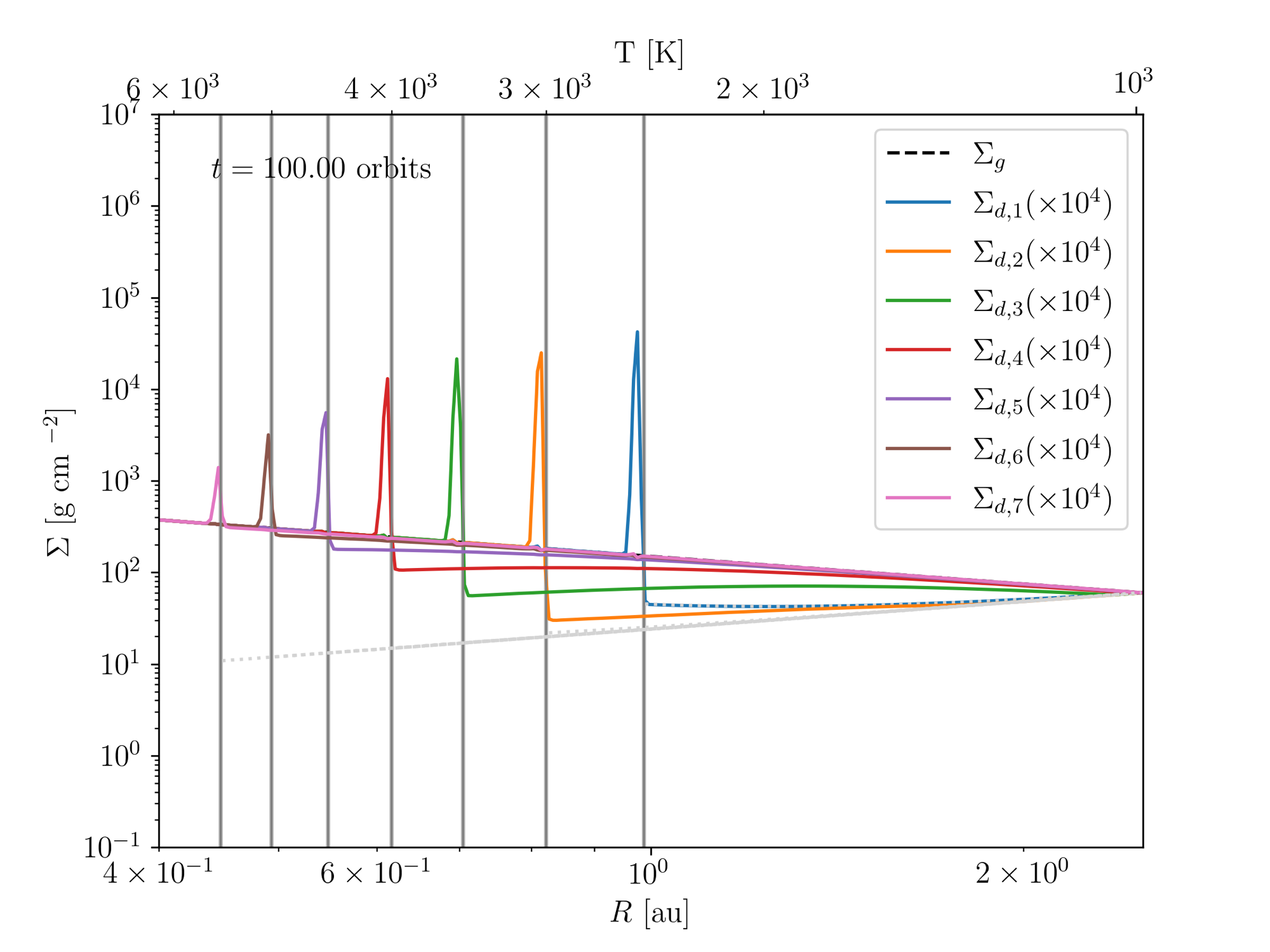}
    \caption{Various dust grain species, with different stopping times and sublimation temperatures, drifting into their respective sublimation fronts (grey regions). At this location, opacity and stopping time both tend to zero, arresting the dust's radial drift and causing it to accumulate. The plotted dust densities $\Sigma_d$ incorporate dust both above and below the sublimation temperature. Dotted grey curves indicate equilibrium dust density profiles expected from the time-independent continuity equation (\ref{eq:continuity_dust}); at low Stokes numbers, these profiles reach a common limit and the curves stack. At $t = 100$ orbits, $\Sigma_{d,1}$ has reached equilibrium outside the sublimation radius, and $\Sigma_{d,2}$ nearly so, but the rest have yet to do so.}
    \label{fig:dust_sublimation_test}
\end{figure}

As dust in a protoplanetary disk drifts radially inward toward the central star, it heats up due to absorption and emission from the background radiation field as well as collisional coupling with the gas. Eventually, it reaches its \textit{sublimation temperature}, at which it undergoes a phase transition from solid to gas; typically, $T_{\rm sub} \approx 150$ K for water ice and 1000-1500 K for various refractory grain species \citep{Duschl1996,Baillie2015}. In the process, standard dust opacity is eliminated, and the sublimated species becomes entrained in the gas, arresting its radial drift and causing it to accumulate at the sublimation radius. Given this ability to naturally concentrate material even in the absence of a pressure bump, the sublimation region is therefore of significant interest as a potential resolution of the radial-drift problem \citep[e.g.,][]{Drazkowska2017,Schoonenberg2017,Wang2025}. Studying this phenomenon requires simultaneously simulating the dynamics and thermodynamics of dust, in both the stiff and non-stiff regimes, making it a valuable showcase of our numerical method.

For this section, we use the disk from Section \ref{sec:radial_drift} as a starting point, with the same number and spacing of cells in the $R$ and $\phi$ directions. The fact that this problem includes radiation, however, breaks scale invariance and requires us to impose physical length ($l_0 = 1$ au), mass ($m_0 = M_*$), and time ($t_0 \equiv \sqrt{l_0^3/Gm_0} = {1 \ \rm y}/2\pi$) scales on this problem. The radial extent of the domain is therefore $R \in [0.4, 2.5] {\rm \ au}$. The surface density of the disk $\Sigma_g(t = 0)$ is given by Equation \ref{eq:sigma_bg}, with a $\Sigma_0 = \SI{32.16}{\gram\per\cm\squared}$. As in the radial-drift setup, the disk's aspect ratio $h \equiv H/R = 0.1$; given the Boltzmann constant $k_B = \SI{1.380649e-16}{\erg\per\kelvin}$ and a mean molecular weight $\mu = 2.3 m_p$, this corresponds to a temperature profile
\begin{equation}
    T_g(R) = \SI{2467}{\kelvin} \left(R/{1 \rm \ au}\right)^{-1}
\end{equation}
We assume the gas follows an ideal equation of state, with $p_g = \rho_g k_BT/\mu$ and an internal energy $E_g = p_g/(\gamma - 1) = c_g\rho_gT$. At the start of the simulation, we set the temperatures of gas, radiation, and all $n_d = 7$ dust species equal to one another. To obtain dust internal energy densities, we assume that the specific heat capacity of each dust species $c_{d,j} = c_g$. For the initial radiation energy density, $E_r \equiv a_r T_r^4$ (with $T_r$ the effective temperature of the radiation), we set $a_r \equiv 4\sigma_{\rm SB}/c$, where the Stefan-Boltzmann constant $\sigma_{\rm SB} = \SI{5.670374e-4}{\erg\per\cm\squared\per\kelvin\tothe{4}}$. We set the reduced speed of light $\hat{c} = 10^{-5}c$.

We approximate dust sublimation within our problem setup by reducing the opacity and stopping time of each species to near-zero at the sublimation temperature. Specifically, for dust species $j$, we use the functional form
\begin{equation}
    \mathcal{S}_j(T) = \left[(1 - \delta_{\rm sub}) (1 + \exp((T - T_{\rm sub,j})/\Delta T_{\rm sub})^{-1} + \delta_{\rm sub}\right] \,,
\end{equation}
where $T_{\rm sub,j} = \SI{2000}{\kelvin} + j \times (\SI{500}{\kelvin})$ is the sublimation temperature of each species, $\Delta T_{\rm sub} = 0.002 T_{\rm sub}$ is a smoothing scale to ensure numerical stability in the iterative calculation, and $\delta_{\rm sub} = 10^{-10}$ is a small factor to prevent singularities in the Jacobian matrix during the implicit step. We can therefore define $\kappa_{d,j} =  \mathcal{S}_j(T_{d,j}) \kappa_0$ and $t_{d,j} = \mathcal{S}_j(T_{d,j}) \left(0.25^j t_0\right)$, where $\kappa_0 = \SI{6.68e2}{cm\tothe{2}\per\gram}$. Such a prescription ensures that sublimated material has no grain opacity, while causing it to become entrained with the gas and follow its bulk motion. This reproduces more comprehensive treatments \citep[e.g.,][]{Wang2023} in the limit that the partial pressures of the sublimated species are small compared to that of the background gas, and can therefore be neglected in the equation of state. The low initial dust-to-gas ratio we choose, $f_{\rm dg,j} = 10^{-4}$ for each species (totaling $f_{\rm dg,j} = 7 \times 10^{-4}$), combined with the higher mean molecular weights of sublimated species (e.g., $\mu_{\rm SiO_2} = 60.08$ and $\mu_{\rm H_2 O} = 16.01$, compared to a $\mu_g = 2.34$ for the gas), helps ensure that this is the case for this test. To stabilize the simulation, we include a nominal viscous and dust-diffusive $\alpha = 10^{-6}$. 

As in the previous full-disk tests in Sections \ref{sec:radial_drift} and \ref{sec:radial_trapping}, we evolve only the dust variables, while freezing all others to their initial condition. Particularly within the sublimation radius, the dynamical and thermal coupling times for gas and dust can be very short. In this highly stiff regime, the operator split between the dust-gas momentum coupling (ordinarily handled within the radiation block when radiation is present) and the pressure, gravitational, and geometric source terms (in the hydrodynamics block) has the potential to introduce error. As such, we incorporate the dust and gas momentum updates into the hydrodynamics block with IMEX1 time integration, as in the purely hydrodynamic tests in Sections \ref{sec:settling_diffusion}, \ref{sec:radial_drift}, and \ref{sec:radial_trapping}), discarding the dust and gas updates obtained in the radiation block.

In Figure \ref{fig:dust_sublimation_test}, we present the results of our sublimation simulation at $t = 100 $ orbits. As predicted, we observe an accumulation of material outside the sublimation radius of each dust species, due to the halt in radial drift caused by the reduction in stopping time; the spatial width of the peak reflects the $\Delta T_{\rm sub}$ over which this reduction is applied. Outside the sublimation radius, the essentially constant rate of radial drift means that a steady-state dust profile can be predicted from the time-independent continuity equation (setting $\partial \Sigma_{g,d}/\partial t = 0$):

\begin{equation}\label{eq:continuity_dust}
    \nabla \cdot (\Sigma_{d,j}\vec{v}_{d,j}) = 0 \rightarrow R\Sigma_{d,j}(\vec{v}_{d,j} \cdot \hat{R}) = R_0 \Sigma_{d,j,0} (\vec{v}_{d,j} \cdot \hat{R})_0
\end{equation}
where we set $(\vec{v}_{d,j} \cdot \hat{R}) = v_{d,r}$ following Equation \ref{eq:radial_drift}, and where the subscript 0 implies values at the outer boundary (where $\Sigma_{d,j}$ and $\vec{v}_{d,j}$ are fixed). At $t = 100 $ orbits, the radial profile for species $j = 1$ has already relaxed to the analytical expectation, and that for $j = 2$ nearly so; for the other dust species, radial drift is slower, so not enough time has passed for their original dust profiles to have been swept away. 

\section{Conclusion}\label{sec:conclusion}
We have designed and implemented a novel numerical method that solves for thermodynamic coupling between gas, dust, and radiation, and in a number of test problems, demonstrate its efficacy and utility in the context of protoplanetary disks. By incorporating multiple species of dust, each with its own self-consistent dynamics (drag and diffusion), our method builds upon its precursor, the three-temperature scheme of \cite{Muley2023}, which only considered a single dust species perfectly coupled to the gas. These changes make it possible to reproduce the spatially varying dust abundances and size distributions created by phenomena such as radial drift \citep[e.g.,][]{Dullemond2018} and dust settling \citep[e.g.,][]{Bae21}. These set the local frequency-integrated opacities and gas-grain thermal accommodation timescales, and in turn the heating, cooling, and background temperature structure of the disk as a whole. 

Monte Carlo radiative transfer (MCRT) tools \citep{Dullemond2012,Whitney2013} are commonly used to convert hydrodynamical simulations into multi-wavelength mock images for comparison with observations \citep[e.g.,][]{Dong2016,Muley2021}, with different species peaking at different wavelengths. With pre-existing tools, the temperatures, velocities, and densities of all relevant species are not readily available from a single simulation, and must be estimated under simplifying assumptions (e.g., dust settling-diffusion balance, dust-gas temperature equality, constant dust-to-gas ratios). With the new method, however, these variables are computed self-consistently within a single simulation, greatly improving the predictive power of mock observations. This would significantly benefit investigations of how large-scale disk substructure, whether driven by planets \citep{Speedie2022}, shadows \citep{Zhang2024_1}, or infalling streamers \citep{Kuznetsova2022}, and their observational appearance in near-infrared scattered light (tracing small grains), millimeter/submillimeter continuum (tracing large grains), and various molecular lines (tracing the gas).

Circumplanetary disks (CPDs) represent another potential area of application for our method. Whether a CPD resembles a rotationally-supported disk or pressure-supported envelope depends on whether thermal relaxation timescales  (relative to compression/expansion timescales) put it in the effectively isothermal or adiabatic regimes, respectively \citep[e.g.,][]{Fung2019}. The opacities and stopping times which mediate thermal relaxation are, in turn, determined by the dynamically evolving size distribution and abundances of dust in the circumplanetary region \citep[][]{Krapp2024b}. From a planet-formation perspective, material delivery to the circumplanetary region is also of great interest, given that individual dust species can be used not only to trace size, but chemical composition (via e.g., the sublimation temperature, specific heat capacity, and material density) as well.

The methods presented in this work are a starting point for a wide range of potential future numerical improvements. Besides collisional coupling, additional processes such as photoelectric heating in the upper disk, and gas line transitions in the inner disk, play a meaningful role in setting disk temperatures \citep{Woitke2009}; the full treatment in thermochemical models such as ProDiMo or DALI could be incorporated into the method by a careful treatment of gas opacities, or by inclusion of additional source terms in the energy update (Equation \ref{eq:energy_update}). A dust vapor module \cite[following ][]{Wang2023} would improve the treatment of dust sublimation, significantly benefiting studies of planet formation, elemental abundance ratios, and the inner disk. For dust dynamics, an improved treatment of dust diffusion \citep[e.g.,][]{Huang2022,Sudarshan2026} or the use of dust turbulent pressure \citep[e.g.,][]{Binkert2023} would be beneficial for treating converging dust flows. One could also generalize the treatment of frictional heating to apply to each dust species separately (that is, write terms $Q_{d,j} \equiv \omega_{d,j} \rho_{d,j} t_{d,j}^{-1} (\vec{v}_{d,j} - \vec{v}_g)^2$, with $\omega_{d,j}$ denoting the fraction of dissipated kinetic energy going to heat up dustspecies $j$) as well as to the ($Q_g = \sum_{j=0}^{n_d} (\omega - \omega_{d,j}) \rho_{d,j} t_{d,j}^{-1} (\vec{v}_{d,j} - \vec{v}_g)^2$). One could also directly incorporating the effects of dynamical coupling into the solution of the Riemann problem \citep[e.g.,][]{Verrier2025}. Higher-order, asymptotically accurate timestepping schemes, such as the variant of ARS(2,2,2) \citep{Ascher1997} investigated by \cite{Krapp2024}, or inspired by the  staggered semi-analytic (SSA) timestepping in \cite{Fung2019b}, would benefit both detailed investigations of disk instabilities and simulations of long-term dust evolution alike. 

\begin{acknowledgements}

We thank Til Birnstiel, Kees Dullemond, Haochang Jiang, Hui Li, Giancarlo Mattia, Alex Mayer, Volker Springel, Peter Woitke, and Shangjia Zhang for useful comments. We also acknowledge helpful and encouraging discussions on scientific applications of this method with Xue-Ning Bai, Mordecai-Mark Mac Low, Nadine Soliman, Zhaohuan Zhu, and other attendees at the Flatiron Workshop on Dusty Turbulence in Protoplanetary Disks, held in February 2026 in New York City. Furthermore, we thank the anonymous referee for a helpful report and suggestions of additional numerical tests, which improved the quality of this paper. Simulations were carried out on the VERA, FREYA, and ORION clusters of the Max Planck Computing and Data Facility (MPCDF), located in Garching bei M\"unchen, Germany. This document makes use of Chentao Yang's \href{https://github.com/yangcht/AA-bibstyle-with-hyperlink}{\texttt{aa\_url}} style file to include references in the bibliography.
\end{acknowledgements}

\bibliography{aa61657-26}{}
\bibliographystyle{aa_url}

\newpage

\onecolumn
\begin{appendix} 
\section{Energy and momentum update}
\subsection{Linearized energy update}\label{sec:rad_matrix_elements}
We linearize Equations \ref{eq:energy_update} in each of the energies, dropping terms involving the derivatives of opacities and dust stopping times:
\begin{subequations}
\begin{equation}\label{eq:e_r}
    \delta X_r/\Delta t = -\delta E_r^{i + 1}/\Delta t - \hat{c} \kappa_{g}^{i+1} \rho_{g} \left(\delta E_r^{i + 1} - 4 a_r T_{g}^{3,i+1} W_{g}^{i+1} \delta E_{g}^{i+1}\right) -\sum_{j = 1}^{n_d} \hat{c} \kappa_{d,j}^{i+1} \rho_{d, j} \left(\delta E_r^{i + 1} - 4 a_r T_{d,j}^{3,i+1} W_{d, j}^{i+1} \delta E_{d, j}^{i+1}\right)
\end{equation}
\begin{equation}\label{eq:e_g}
    \delta X_g/\Delta t = -\delta E_g^{i+1}/\Delta t + c \kappa_g^{i+1} \rho_g (\delta E_r^{i+1} - 4 a_r T_g^{3,i+1}W_g^{i+1}\delta E_{g}^{i+1}) + \sum_{j = 1}^{n_d} \eta_{d,j}^{i+1} t_{d, j}^{-1, i+1}  \rho_{d,j} \left(2 k_B W_{d, j}^{i+1} \delta E_{d, j}^{i+1} - 2 k_B W_g^{i+1}\delta E_{g}^{i+1}\right)
\end{equation}
\begin{equation}\label{eq:e_d_j}
    \delta X_{d,j}/\Delta t = -\delta E_{d,j}^{i+1}/\Delta t + c \kappa_{d,j}^{i+1} \rho_{d,j} (\delta E_r^{i+1} - 4 a_r T_{d,j}^{3,i+1}W_{d,j}^{i+1}\delta E_{d,j}^{i+1}) - \eta_{d,j}^{i+1} t_{d, j}^{-1, i+1}  \rho_{d,j} \left(2 k_B W_{d, j}^{i+1} \delta E_{d, j}^{i+1} - 2 k_B W_g^{i+1}\delta E_{g}^{i+1}\right)
\end{equation}
\end{subequations}
Where $W = \bar{W}/ \rho$ represents the inverse heat capacity at constant volume. In general, $W$ can be any function of the primitive variables, but for all tests presented, we use an ideal equation of state for the gas with a constant adiabatic index $\gamma$, as well as a constant heat capacity for the dust, implying a constant $W$.

As described in Section \ref{sec:iterative_procedure}, energy exchange between gas, dust, and radiation is more efficiently computed using a reduced $2 \times 2$ system $\delta \vec{X}' = \mathbf{\Tilde{J}}'\delta \vec{E'}^{i+1}$, rather than the full system $\delta \vec{X} = \mathbf{\Tilde{J}}\delta \vec{E}^{i+1}$. For this problem, we can write a reduced energy vector $\delta \vec{E'}^{i+1} = \left(\delta E_r^{i+1}, \delta E_g^{i+1}\right)$. To compute the dust energy update, we rearrange Equation \ref{eq:e_d_j} to yield the following:
\begin{equation}\label{eq:e_d_j_reduced}
\begin{split}
    \delta E_{d,j}^{i+1} = \frac{\left(c \kappa_{d,j}^{i+1} \rho_{d,j} \delta E_r^{i+1} + \eta_{d,j}^{i+1} \rho_{d,j} t_{d, j}^{-1, i+1} 2 k_B W_g^{i+1}\delta E_{g}^{i+1}\right)\Delta t - \delta X_{d,j}}{1 + \left(4 c \kappa_{d,j}^{i+1} \rho_{d,j} a_r T_{d,j}^{3,i+1}W_{d,j}^{i+1} + \eta_{d,j}^{i+1} \rho_{d,j} t_{d, j}^{-1, i+1} 2  k_B W_{d, j}^{i+1}\right) \Delta t} \;
    \end{split}
\end{equation}

Substituting Equations \ref{eq:e_d_j_reduced} into Equation \ref{eq:e_r} and \ref{eq:e_g} yields the following expressions for $\delta X_r$ and $\delta X_g$ purely in terms of gas and radiation variables:
\begin{subequations}
\begin{equation}\label{eq:e_r_reduced}
\begin{split}
    \delta X_r/\Delta t = &-\delta E_r^{i + 1}/\Delta t - \hat{c} \kappa_{g}^{i+1} \rho_{g} \left(\delta E_r^{i + 1} - 4 a_r T_{g}^{3,i+1} W_{g}^{i+1} \delta E_{g}^{i+1}\right) 
    \\ &-\sum_{j = 1}^{n_d} \hat{c} \kappa_{d,j}^{i+1} \rho_{d, j} \left(\delta E_r^{i + 1} -  4 a_r T_{d,j}^{3,i+1} W_{d, j}^{i+1} \left[\frac{\left(c \kappa_{d,j}^{i+1} \rho_{d,j} \delta E_r^{i+1} + \eta_{d,j}^{i+1} t_{d, j}^{-1, i+1} 2 k_B W_g^{i+1}\delta E_{g}^{i+1}\right)\Delta t  - \delta X_{d,j}}{1 + \left(4 c \kappa_{d,j}^{i+1} \rho_{d,j} a_r T_{d,j}^{3,i+1}W_{d,j}^{i+1} + \eta_{d,j}^{i+1} \rho_{d,j} t_{d, j}^{-1, i+1}2 k_B W_{d, j}^{i+1}\right) \Delta t}\right]\right)%\\\\
\end{split}
\end{equation}
\begin{equation}\label{eq:e_g_reduced}
\begin{split}
    \delta X_g/\Delta t = &-\delta E_g^{i+1}/\Delta t + c \kappa_g^{i+1} \rho_g (\delta E_r^{i+1} - 4 a_r T_g^{3,i+1}W_g^{i+1}\delta E_{g}^{i+1}) \\
    &+ \sum_{j = 1}^{n_d} \eta_{d,j}^{i+1} \rho_{d,j} 
t_{d, j}^{-1, i+1} \left(2 k_B W_{d, j}^{i+1} \left[\frac{\left(c \kappa_{d,j}^{i+1} \rho_{d,j} \delta E_r^{i+1} + \eta_{d,j}^{i+1}  \rho_{d,j} t_{d, j}^{-1, i+1} 2 k_B W_g^{i+1}\delta E_{g}^{i+1}\right)\Delta t - \delta X_{d,j}}{1 + \left(4 c \kappa_{d,j}^{i+1} \rho_{d,j} a_r T_{d,j}^{3,i+1}W_{d,j}^{i+1} + \eta_{d,j}^{i+1}  \rho_{d,j} t_{d, j}^{-1, i+1}2 k_B W_{d, j}^{i+1}\right) \Delta t}\right] - 2 k_B W_g^{i+1}\delta E_{g}^{i+1}\right)%\\\\
\end{split}
\end{equation}
\end{subequations}
We then move all expressions involving $\delta X_{d,j}$ from the right-hand side of the system to the left, and assign these quantities to $\vec{X}'\equiv \left(\delta X_r', \delta X_g'\right)$. The resulting matrix and vector elements are as follows:

\begin{subequations}
\begin{equation}
    \delta X_r' = \delta X_r +\sum_{j = 1}^{n_d}  \delta X_{d,j} \left[\frac{4 a_r T_{d,j}^{3,i+1} W_{d, j}^{i+1} \hat{c} \kappa_{d,j}^{i+1} \rho_{d, j} \Delta t}{1 + \left(4 c \kappa_{d,j}^{i+1} \rho_{d,j} a_r T_{d,j}^{3,i+1}W_{d,j}^{i+1} + \eta_{d,j}^{i+1}  \rho_{d,j} t_{d, j}^{-1, i+1}2 k_B W_{d, j}^{i+1}\right) \Delta t}\right]
\end{equation}
\begin{equation}
    \delta X_g' = \delta X_g + \sum_{j = 1}^{n_d}  \delta X_{d,j} \left[\frac{2 k_B W_{d, j}^{i+1} \eta_{d,j}^{i+1}  \rho_{d,j} t_{d, j}^{-1, i+1} \Delta t}{1 + \left(4 c \kappa_{d,j}^{i+1} \rho_{d,j} a_r T_{d,j}^{3,i+1}W_{d,j}^{i+1} + \eta_{d,j}^{i+1}  \rho_{d,j} t_{d, j}^{-1, i+1}2 k_B W_{d, j}^{i+1}\right) \Delta t}\right]
\end{equation}
\end{subequations}
\begin{subequations}
\begin{equation}
\begin{split}
    \Tilde{J}'_{rr} = &-\left\{1 + \hat{c} \kappa_{g}^{i+1} \rho_{g} \Delta t + \sum_{j = 1}^{n_d} \hat{c} \kappa_{d,j}^{i+1} \rho_{d, j} \Delta t \left[\frac{1 + 2 k_B W_{d, j}^{i+1} \eta_{d,j}^{i+1}  \rho_{d,j} t_{d, j}^{-1, i+1}\Delta t}{1 + \left(4 a_r T_{d,j}^{3,i+1}W_{d,j}^{i+1} c \kappa_{d,j}^{i+1} \rho_{d,j} 
    + \eta_{d,j}^{i+1} t_{d, j}^{-1, i+1}2 k_B W_{d, j}^{i+1}\right) \Delta t}\right]\right\}\\
    = &-\left\{1 + \frac{\hat{c}}{c} \Tilde{J}'_{gr} + \sum_{j = 1}^{n_d} \left[\frac{\hat{c} \kappa_{d,j}^{i+1} \rho_{d, j} \Delta t}{1 + \left(4 a_r T_{d,j}^{3,i+1}W_{d,j}^{i+1} c \kappa_{d,j}^{i+1} \rho_{d,j} 
    + \eta_{d,j}^{i+1}  \rho_{d,j} t_{d, j}^{-1, i+1}2 k_B W_{d, j}^{i+1}\right) \Delta t}\right]\right\}
\end{split}
\end{equation}
\begin{equation}
\begin{split}
    \Tilde{J}'_{rg} = & \left\{ \left(4 a_r T_{g}^{3,i+1} W_{g}^{i+1}\right) \hat{c} \kappa_{g}^{i+1} \rho_{g} \Delta t \right.\\
    & \left. + \sum_{j = 1}^{n_d} \eta_{d,j}^{i+1}  \rho_{d,j} t_{d, j}^{-1, i+1} 2 k_B W_g^{i+1}\Delta t \left[\frac{\left(4 a_r T_{d,j}^{3,i+1} W_{d, j}^{i+1}\right) \hat{c} \kappa_{d,j}^{i+1} \rho_{d, j} \Delta t}{1 + \left(4 c \kappa_{d,j}^{i+1} \rho_{d,j} a_r T_{d,j}^{3,i+1}W_{d,j}^{i+1} + \eta_{d,j}^{i+1}  \rho_{d,j} t_{d, j}^{-1, i+1}2 k_B W_{d, j}^{i+1}\right) \Delta t}\right]\right\}
\end{split}
\end{equation}
\begin{equation}
     \Tilde{J}'_{gr} = \left\{c \kappa_g^{i+1} \rho_g \Delta t  + \sum_{j = 1}^{n_d}c \kappa_{d,j}^{i+1} \rho_{d,j} \Delta t \left[\frac{ 2 k_B W_{d, j}^{i+1} \eta_{d,j}^{i+1}  \rho_{d,j} t_{d, j}^{-1, i+1} \Delta t}{1 + \left(4 c \kappa_{d,j}^{i+1} \rho_{d,j} a_r T_{d,j}^{3,i+1}W_{d,j}^{i+1} + \eta_{d,j}^{i+1}  \rho_{d,j} t_{d, j}^{-1, i+1}2 k_B W_{d, j}^{i+1}\right) \Delta t}\right]\right\}
\end{equation}
\begin{equation}
\begin{split}
    \Tilde{J}'_{gg} = &-\left\{1 + 4 a_r T_g^{3,i+1}W_g^{i+1} c \kappa_g^{i+1} \rho_g \Delta t \right.\\
    &\left. + \sum_{j = 1}^{n_d} \eta_{d,j}^{i+1}  \rho_{d,j} t_{d, j}^{-1, i+1} 2 k_B W_{g}^{i+1} \Delta t \left[\frac{1 + 4 c \kappa_{d,j}^{i+1} \rho_{d,j} a_r T_{d,j}^{3,i+1}W_{d,j}^{i+1} \Delta t}{1 + \left(4 c \kappa_{d,j}^{i+1} \rho_{d,j} a_r T_{d,j}^{3,i+1}W_{d,j}^{i+1} + \eta_{d,j}^{i+1}  \rho_{d,j} t_{d, j}^{-1, i+1}2 k_B W_{d, j}^{i+1}\right) \Delta t}\right]\right\}\\
     = &-\left\{1 + \frac{c}{\hat{c}}\Tilde{J}'_{rg} + \sum_{j = 1}^{n_d}  \left[\frac{\eta_{d,j}^{i+1}  \rho_{d,j} t_{d, j}^{-1, i+1} 2 k_B W_{g}^{i+1} \Delta t}{1 + \left(4 c \kappa_{d,j}^{i+1} \rho_{d,j} a_r T_{d,j}^{3,i+1}W_{d,j}^{i+1} + \eta_{d,j}^{i+1}  \rho_{d,j}
 t_{d, j}^{-1, i+1}2 k_B W_{d, j}^{i+1}\right) \Delta t}\right]\right\}
\end{split}
\end{equation}
\end{subequations}
Each element of $\mathbf{\Tilde{J}}'$ and $\delta \vec{X}'$ contains terms corresponding to the $n_d$ dust species, plus the single gas species (as well as an additional identity term along the diagonals). These terms share broadly similar functional forms, which can be used to reduce the total number of function evaluations. The reduced system can be solved for each element of $\delta \vec{E'}^{i+1}$, after which we use Equation \ref{eq:e_d_j_reduced} to compute each of the dust energies at each iteration.
\subsection{Linearized momentum update}\label{sec:momentum_matrix_elements}

As before, we linearize the system of equations for momentum (Equation \ref{eq:momentum_exchange}) and rearrange so the residual is on the left:
\begin{subequations}
\begin{equation}
    \delta \vec{R}_r/\delta t = -\left[\frac{1}{\delta t} + \hat{c}\left(\rho_g \chi_g^{i+1} + \sum_{j=1}^{n_d}\rho_{d,j}\chi_{d,j}^{i+1}\right)\right]\delta \vec{F}_r^{i+1}
\end{equation}
\begin{equation}
    \delta \vec{R}_g/\delta t = -\delta \vec{p}_g^{i+1}/\delta t + \rho_g\chi_g^{i+1}\delta \vec{F}_r^{i+1} + \sum_{j=1}^{n_d} \rho_{d,j}t_{d,j}^{-1,i+1}\left(\frac{\delta \vec{p}_{d,j}^{i+1}}{\rho_{d,j}} - \frac{\delta \vec{p}_{g}^{i+1}}{\rho_g}\right)
\end{equation}
\begin{equation}
    \delta \vec{R}_{d,j}/\delta t = -\delta \vec{p}_{d,j}^{i+1}/\delta t + \rho_{d,j}\chi_{d,j}^{i+1}\delta \vec{F}_r^{i+1} - \rho_{d,j}t_{d,j}^{-1,i+1}\left(\frac{\delta \vec{p}_{d,j}^{i+1}}{\rho_{d,j}} - \frac{\delta \vec{p}_{g}^{i+1}}{\rho_g}\right)
\end{equation}
\begin{equation}
\end{equation}
\end{subequations}
Because the flux does not explicitly depend on the other momentum variables, there is an analytical solution for which $\vec{R}_r$ is zero. Inserting this solution into the original equations immediately allows any terms involving $\delta \vec{R}_r$ or $\delta \vec{F}_r^{i+1}$ to be dropped.
\begin{subequations}
\begin{equation}
    \delta \vec{R}_g
    = -\delta \vec{p}_g^{i+1}\left[1 + \delta t \sum_{j=1}^{n_d} \frac{\rho_{d,j}t_{d,j}^{-1,i+1}}{\rho_g}\right] + \left[\delta t \sum_{j=1}^{n_d} t_{d,j}^{-1,i+1} \delta \vec{p}_{d,j}^{i+1}\right]
\end{equation}
\begin{equation}
    \delta \vec{R}_{d,j}
    = -\delta \vec{p}_{d,j}^{i+1}\left[1 + t_{d,j}^{-1,i+1} \delta t\right] + \delta \vec{p}_{g}^{i+1} \frac{\rho_{d,j}}{\rho_g} t_{d,j}^{-1}\delta t
\end{equation}
\end{subequations}
Some algebra then yields that
\begin{subequations}
\begin{equation}\label{eq:fast_linear}
    \delta \vec{R}_g + \sum_{j=1}^{n_d} \delta \vec{R}_{d,j} \frac{t_{d,j}^{-1,i+1} \delta t}{1 + t_{d,j}^{-1,i+1} \delta t} = -\delta \vec{p}_g^{i+1}\left[1 + \sum_{j=1}^{n_d}   \frac{\rho_{d,j}}{\rho_g} \left(\frac{t_{d,j}^{-1,i+1} \delta t}{1 + t_{d,j}^{-1,i+1} \delta t}\right)\right]
\end{equation}
    \begin{equation}\label{eq:dust_calc}
    \delta \vec{p}_{d,j}^{i+1} = \frac{\delta \vec{p}_{g}^{i+1} (\rho_{d,j}/\rho_{g})t_{d,j}^{-1,i+1} \delta t - \delta \vec{R}_{d,j}}{1 + t_{d,j}^{-1,i+1} \delta t}
\end{equation}
\end{subequations}

The $\delta \vec{p}_{d,j}^{i+1}$ and $\delta \vec{p}_{g}^{i+1}$ computed from Equations \ref{eq:fast_linear} and \ref{eq:dust_calc} bear similarities to the fast dust momentum calculation presented in \citep{BenitezLlambay2019}. Indeed, applying that work's assumption of linear drag and no radiative flux, they can be used to compute exact solutions for momentum change over a timestep, as detailed in Appendix \ref{sec:krapp_comparison}. In our simulations, these assumptions do not hold in general, because opacities and dust-gas stopping times may depend on temperature \citep[and, for supersonic drag, on the velocity; ][]{Hopkins2016}, and need to be evaluated iteratively in alternation with energy exchange.
\subsection{Conservation laws}\label{sec:conservation_laws_appendix}

\subsubsection{Energy}
Summing together the original equations, for any assumed $\vec{E}^{i+1,*k}$, and multiplying by $\delta t$ yields 
\begin{equation}\label{eq:total_sum}
    (c/\hat{c})(E_r^{i+1,*k} - E_r^{i}) + (E_g^{i+1,*k} - E_g^{i}) + \sum_{j=1}^{n_d} (E_{d,j}^{i+1,*k} - E_{d,j}^{i}) = -\left[(c/\hat{c}) X_r^{*k} + X_g^{*k} + \sum_{j=1}^{n_d}  X_{d,j}^{*k}\right] + \left[Q_g^{i+1}
    + S_g + \sum_{j=1}^{n_d}  S_{d,j}\right] \delta t
\end{equation}
In other words, the change in the total energy of the system (modified to account for the reduced speed of light $\hat{c}$) is equal to the contribution from frictional heating $Q_g^{i+1}$ (computed from the momentum block) and stellar irradiation $S$ onto each of the species. The (again, modified) sum of energy residuals is a measure of the error in total energy accrued during an iteration.

Linearizing the above (or equivalently, taking the sum of the linearized individual equations) yields that 
\begin{equation}\label{eq:linearized_sum}
    (c/\hat{c})\delta E_r^{i+1,*k} + \delta E_g^{i+1,*k} + \sum_{j=1}^{n_d} \delta E_{d,j}^{i+1,*k} = -\left[(c/\hat{c}) \delta X_r^{*k} + \delta X_g^{*k} + \sum_{j=1}^{n_d}  \delta X_{d,j}^{*k}\right]
\end{equation}
For the Newton-Raphson technique, as mentioned earlier, we set $\delta \vec{X}^{*k} = -\vec{X}^{*k}$, as mentioned earlier. Making this substitution and summing Equations \ref{eq:total_sum} and \ref{eq:linearized_sum} yields
\begin{equation}\label{eq:conservation_sum}
    (c/\hat{c})(E_r^{i+1,*k+1} - E_r^{i}) + (E_g^{i+1,*k+1} - E_g^{i}) + \sum_{j=1}^{n_d} (E_{d,j}^{i+1,*k+1} - E_{d,j}^{i}) = \left[Q_g^{i+1} %+ Q_r^{i+1}
    + S_g + \sum_{j=1}^{n_d}  S_{d,j}\right] \delta t
\end{equation}
where the residuals cancel, and terms of the form $E_x^{i+1,*k} + \delta E_x^{i+1,*k}$ have been identified with the energy value at the next iteration, $E_x^{i+1,*k+1}$. Finally, replacing $k \rightarrow k + 1$ in Equation \ref{eq:total_sum}, and subtracting Equation \ref{eq:conservation_sum} from it, yields that
\begin{equation}
    0 = -\left[(c/\hat{c}) X_r^{*k+1} + X_g^{*k+1} + \sum_{j=1}^{n_d}  X_{d,j}^{*k+1}\right]
\end{equation}
Intuitively, this implies that each Newton-Raphson iteration enforces manifest conservation of total energy, up to the desired source terms and the $c/\hat{c}$ factor applied to the radiation energy, regardless of whether the initial guess is energy-conserving. The same conservative property is also present in the methods of \cite{MelonFuksman2021} and \cite{Muley2023}. This does not imply convergence of the method in a single iteration, as the individual elements of the residual vector need not be not linear in the energies and must be re-evaluated at each iteration.
\subsubsection{Momentum}
We can similarly obtain the momentum conservation law, using the $\vec{p}^{i+1,*k}$ to denote the entire list of momenta/fluxes (each of whose components is itself a vector in space):
\begin{equation}
    (1/\hat{c})(\vec{F}_r^{i+1,k*} - \vec{F}_r^{i}) + (\vec{p}_g^{i+1,k*} - \vec{p}_g^{i}) + \sum_{j=1}^{n_d} (\vec{p}_{d,j}^{i+1,k*} - \vec{p}_{d,j}^{i}) = -\left[\vec{R}_r^{*k} + \vec{R}_g^{*k} + \sum_{j=0}^{n_d} \vec{R}_{d,j}^{*k} \right] \,.
\end{equation}
Summing over the linearized problem yields
\begin{equation}
    (1/\hat{c})\delta \vec{F}_r^{i+1,*k} + \delta \vec{p}_g^{i+1,*k} + \sum_{j=1}^{n_d} \delta \vec{p}_{d,j}^{i+1,*k} = -\left[\delta \vec{R}_r^{*k} + \delta \vec{R}_g^{*k} + \sum_{j=0}^{n_d} \delta \vec{R}_{d,j}^{*k} \right] \,.
\end{equation}
We apply again the assumption in the Newton-Raphson method, that $\delta \vec{R}^{*k} = -\vec{R}^{*k}$ (i.e., we want to solve for a zero residual in the linearized problem). Identifying $\vec{p}_x^{i+1,*k+1} = \vec{p}_x^{i+1,*k} + \vec{p}_x^{i+1,*k}$ yields that
\begin{equation}
    (1/\hat{c})(\vec{F}_r^{i+1,*k+1} - \vec{F}_r^{i}) + (\vec{p}_g^{i+1,*k+1} - \vec{p}_g^{i}) + \sum_{j=1}^{n_d} (\vec{p}_{d,j}^{i+1,*k+1} - \vec{p}_{d,j}^{i}) = 0 \,,
\end{equation}
which implies conservation of momentum as well, up to a factor of $(c/\hat{c})$ applied to the fluxes, in our implicit step.

\section{Comparison to \cite{BenitezLlambay2019}}\label{sec:krapp_comparison}

Using the initial guess that $\vec{p}^{i+1,*0} = \vec{p}^{i}$,  setting radiative fluxes to zero, and assuming constant stopping time (i. e., a linear drag law), we obtain that
\begin{equation}
    \vec{R}_g = \sum_{j=1}^{n_d} \rho_{d,j} t_{d,j}^{-1} \Delta t \left(\vec{v}_{d,j}^{i} - \vec{v}_g^{i}\right) \,,
\end{equation}
\begin{equation}
    \vec{R}_{d,j} = -\rho_{d,j} t_{d,j}^{-1} \Delta t \left(\vec{v}_{d,j}^{i} - \vec{v}_g^{i}\right) \,.
\end{equation}
Substituting $\vec{R} = -\delta \vec{R}$ into equation (13a) yields that
\begin{equation}\label{eq:krapp_substituted_pg}
     \left(\vec{v}_g^{i+1} - \vec{v}_g^{i}\right) = \left[\sum_{j=1}^{n_d}\frac{\rho_{d,j}}{\rho_g} \left(\vec{v}_{d,j}^{i} - \vec{v}_g^{i}\right) \left(\frac{t_{d,j}^{-1} \Delta t}{1 + t_{d,j}^{-1} \Delta t}\right)\right]\left[1 + \sum_{j=1}^{n_d}   \frac{\rho_{d,j}}{\rho_g} \left(\frac{t_{d,j}^{-1} \Delta t}{1 + t_{d,j}^{-1} \Delta t}\right)\right]^{-1} \,,
\end{equation}
    \begin{equation}\label{eq:krapp_substituted_pd}
    \left(\vec{v}_{d,j}^{i+1} - \vec{v}_{d,j}^{i}\right) = \frac{\left[(\vec{v}_g^{i+1} - \vec{v}_g^{i}) - (\vec{v}_{d,j}^{i} - \vec{v}_g^{i})\right]  t_{d,j}^{-1} \Delta t}{1 + t_{d,j}^{-1} \Delta t} \,.
\end{equation}

\section{Geometric source terms for diffusion}\label{sec:geom_source_terms}
The advection of bulk momentum by the diffusion term $\vec{S}_{m,d,j} = -\nabla \cdot \left(\vec{j}_{{\rm diff},j} \vec{v}_{d,j}\right) = -\nabla \cdot \left(\rho_{d,j} \vec{v}_{{\rm diff},j} \vec{v}_{d,j}\right)$ takes the form of a tensor divergence (for a tensor $\mathbf{S}_{m,d,j} \equiv \rho_{d,j} \vec{v}_{{\rm diff},j} \vec{v}_{d,j}$), analogously to the advection of momentum in the conservative Euler equations. This tensorial divergence can be decomposed into a rate of transport of each vector component, and the rate of change in the coordinate basis vectors due to transport (the so-called ``geometric source terms''). In what follows, we present explicit functional forms for the source terms.

\subsection{Spherical coordinates}
In spherical coordinates, the conservative and geometric source terms take the following functional forms:

\begin{subequations}\label{eq:geom_source_terms_spherical}
\begin{equation}
\begin{split}
    \vec{S}_{m,d,j} \cdot \hat{r} &= -\nabla \cdot \left[\rho_{d,j} (\vec{v}_{d,j} \cdot \hat{r}) \vec{v}_{\rm diff,j}\right] + \rho_{d,j} r^{-1}\left[(\vec{v}_{\rm diff,j} \cdot \hat{\theta})(\vec{v}_{d,j} \cdot \hat{\theta}) + (\vec{v}_{\rm diff,j} \cdot \hat{\phi})(\vec{v}_{d,j} \cdot \hat{\phi})\right]
\end{split}
\end{equation}
\begin{equation}
\begin{split}
    \vec{S}_{m,d,j} \cdot \hat{\theta} &= -\nabla \cdot \left[\rho_{d,j} (\vec{v}_{d,j} \cdot \hat{\theta}) \vec{v}_{\rm diff,j}\right] - \rho_{d,j}r^{-1}\left[(\vec{v}_{\rm diff,j} \cdot \hat{\theta})(\vec{v}_{d,j} \cdot \hat{r}) - \cot \theta (\vec{v}_{\rm diff,j} \cdot \hat{\phi})(\vec{v}_{d,j} \cdot \hat{\phi})\right]
\end{split}
\end{equation}
\begin{equation}
\begin{split}
    \vec{S}_{m,d,j} \cdot \hat{\phi} &= -\nabla \cdot \left[\rho_{d,j} (\vec{v}_{d,j} \cdot \hat{\phi}) \vec{v}_{\rm diff,j}\right] - \rho_{d,j} r^{-1}\left[(\vec{v}_{\rm diff,j} \cdot \hat{\phi})(\vec{v}_{d,j} \cdot \hat{r}) + \cot \theta (\vec{v}_{\rm diff,j} \cdot \hat{\phi})(\vec{v}_{d,j} \cdot \hat{\theta})\right] \,.
\end{split}
\end{equation}
\end{subequations}

In our implementation, the divergence term is moved to the left-hand side and is incorporated into the flux computed at cell faces, while the geometric term remains on the right and is computed as a cell-centered quantity. In the $\hat{\phi}$ direction, however, \texttt{PLUTO} uses an ``augmented divergence'' $\nabla^r$, which effectively converts the equation from a conservation law for linear momentum to one for angular momentum:
\begin{equation}
\begin{split}
    \nabla^r \cdot \vec{F}_k &\equiv (r \sin \theta)^{-1} \nabla \cdot (r \sin \theta \vec{F}_k ) \\
    &= \nabla \cdot  \vec{F}_k + (r \sin \theta)^{-1}\vec{F}_k \cdot \nabla(r\sin\theta)\\
    &= \nabla \cdot \vec{F}_k  + r^{-1}(\vec{F}_k \cdot{\hat{r}}) + r^{-1}(\vec{F}_k \cdot{\hat{\theta}})\cot \theta \,.
\end{split}
\end{equation}
For a symmetric momentum tensor, as is the case in the standard Euler equations, this formulation absorbs all $\hat{\phi}$ geometric source terms, implying that angular momentum is conserved. In our formulation, however, the dust diffusion tensor $\mathbf{S}_{m,d,j}$ is asymmetric, with the antisymmetric part of the tensor contributing to non-ideal source terms and generating angular momentum:

\begin{equation}\label{eq:augmented_div_diffusion}
\begin{split}
    \vec{S}_{m,d,j} \cdot \hat{\phi} = &-\nabla^r \cdot \left[\rho_{d,j} (\vec{v}_{d,j} \cdot \hat{\phi}) \vec{v}_{\rm diff,j}\right]\\ 
    & -\rho_{d,j} r^{-1}\left[(\vec{v}_{\rm diff,j} \cdot \hat{\phi})(\vec{v}_{d,j} \cdot \hat{r}) + \cot \theta (\vec{v}_{\rm diff,j} \cdot \hat{\phi})(\vec{v}_{d,j} \cdot \hat{\theta})\right]\\ 
    & +\rho_{d,j} r^{-1}\left[(\vec{v}_{\rm diff,j} \cdot \hat{r})(\vec{v}_{d,j} \cdot \hat{\phi}) + \cot \theta (\vec{v}_{\rm diff,j} \cdot \hat{\theta})(\vec{v}_{d,j} \cdot \hat{\phi})\right]
\end{split}
\end{equation}
More detailed formulations of dust diffusion \citep[e.g.,][]{Huang2022,Binkert2023} aim to remedy this issue by employing a symmetrized diffusive momentum tensor, $\mathbf{\bar{S}}_{m,d,j} \equiv \mathbf{S}_{m,d,j} + \mathbf{S}^{\top}_{m,d,j} = \rho_{d,j} \vec{v}_{{\rm diff},j} \vec{v}_{d,j} + \rho_{d,j} \vec{v}_{d,j}\vec{v}_{{\rm diff},j}$, where $\vec{S}_{m,d,j} \equiv \nabla \cdot \mathbf{\bar{S}}_{m,d,j}$. The components of $\nabla \cdot \mathbf{S}^{\top}_{m,d,j}$, and in turn of the symmetrized $\nabla \cdot \mathbf{\bar{S}}_{m,d,j}$ can be found simply by exchanging $\vec{v}_{d,j}$ with $\vec{v}_{\rm diff,j}$ in each of Equations \ref{eq:geom_source_terms_spherical} and \ref{eq:augmented_div_diffusion}; doing so demonstrates that the $\hat{\phi}$ geometric source terms disappear when the symmetrized formulation is used. Nevertheless, using this tensor leads to additional complications at the implementation level, and given that the velocities associated with dust diffusion ($c_s \rm{St}$) are much lower than those of the bulk dust motion ($v_k$), we defer implementation of this symmetrized tensor to a dedicated work.
\subsection{Cylindrical-polar coordinates}
In a cylindrical-polar coordinate system, the expressions read:
\begin{subequations}\label{eq:geom_source_terms_cylindrical}
\begin{equation}
\begin{split}
    \vec{S}_{m,d,j} \cdot \hat{R} &= -\nabla \cdot \left[\rho_{d,j} (\vec{v}_{d,j} \cdot \hat{R}) \vec{v}_{\rm diff,j}\right] + \rho_{d,j} R^{-1}\left[(\vec{v}_{\rm diff,j} \cdot \hat{\phi})(\vec{v}_{d,j} \cdot \hat{\phi})\right]
\end{split}
\end{equation}
\begin{equation}
\begin{split}
    \vec{S}_{m,d,j} \cdot \hat{\phi} &= -\nabla \cdot \left[\rho_{d,j} (\vec{v}_{d,j} \cdot \hat{\phi}) \vec{v}_{\rm diff,j}\right] - \rho_{d,j}R^{-1}\left[(\vec{v}_{\rm diff,j} \cdot \hat{\phi})(\vec{v}_{d,j} \cdot \hat{R})\right]
\end{split}
\end{equation}
\begin{equation}
\begin{split}
    \vec{S}_{m,d,j} \cdot \hat{z} &= -\nabla \cdot \left[\rho_{d,j} (\vec{v}_{d,j} \cdot \hat{z}) \vec{v}_{\rm diff,j}\right] \,.
\end{split}
\end{equation}
\end{subequations}
Analogously to the spherical case, one can define an augmented divergence $\nabla^R \cdot \vec{F}_k \equiv R^{-1}\nabla \cdot (R\vec{F}_k)$ to express the conservation law for the $\hat{\phi}$-component in terms of angular momentum:
\begin{equation}
\begin{split}
    \vec{S}_{m,d,j} \cdot \hat{\phi} &= -\nabla^R \cdot \left[\rho_{d,j} (\vec{v}_{d,j} \cdot \hat{\phi}) \vec{v}_{\rm diff,j}\right] - \rho_{d,j}R^{-1}\left[(\vec{v}_{\rm diff,j} \cdot \hat{\phi})(\vec{v}_{d,j} \cdot \hat{R})\right] + \rho_{d,j}R^{-1}\left[(\vec{v}_{\rm diff,j} \cdot \hat{R})(\vec{v}_{d,j} \cdot \hat{\phi})\right]
\end{split}
\end{equation}
where the functional form again makes clear that it is the antisymmetric component of the diffusion-velocity tensor that gives rise to non-conservation of angular momentum.
\section{Test problem derivations}
\subsection{Irradiation heating term}\label{sec:irr_appendix}
In the case where there are multiple absorbing dust species, the monochromatic radiative transfer equation (RTE) is given by
\begin{equation}\label{eq:formal_rt_eqn}
    \frac{\partial I_{b}}{\partial t} + c \nabla \cdot (I_{b}\hat{n}) = \sum_{l=1}^{n_d}c \kappa_{d,l,b}\rho_{d,l}(S_{b}(T_{d,l})-I_{b}) \,,
\end{equation}
where $I_{b}$ is the specific intensity, $\hat{n}$ the direction in question, and $S_{b}$ the source function corresponding to frequency band $b$. The zeroth moment of this equation,
\begin{equation}\label{eq:formal_rt_eqn_mom1}
    \frac{\partial E_{b}}{\partial t} + c \nabla \cdot \vec{F}_{b} = \sum_{l=1}^{n_d}c \kappa_{d,l,b}\rho_{d,l}(B_{b}(T_{d,j})-E_{b}) \,,
\end{equation}
gives the rate of change for the radiation energy density in each band, whereas 
\begin{equation}\label{eq:dust_energy_change}
    S_{\rm irr,d,j} = -\sum_{b 
    %\subseteq b_{\rm irr}
    }S_{\rm irr,d,j,b} = -\sum_{b %\subseteq b_{\rm irr}
    } c \kappa_{d,j,b}\rho_{d,j}(B_{b}(T_{d,j})-E_{b})
\end{equation}
gives the rate at which species $j$ is heated by its interaction with all irradiation bands $b$. Dropping the time dependence in Equation \ref{eq:formal_rt_eqn_mom1} and combining it with \ref{eq:dust_energy_change} yields the formulation
\begin{equation}\label{eq:rt_reformulated}
    c\nabla \cdot \vec{F}_{b} + \sum_{l=1}^{n_d}S_{\rm irr,d,l,b} = 0 \rightarrow S_{\rm irr,d,j,b} = -\frac{S_{\rm irr,d,j,b}}{\sum_{l=1}^{n_d}S_{\rm irr,d,l,b}} \nabla \cdot c\vec{F}_{b}
\end{equation}

In protoplanetary-disk applications, the temperature of the thermally reprocessed radiation ($B_b$) is typically at least an order of mangitude lower than the effective temperature of the incident stellar irradiation ($E_b$). As such, the frequency content of these spectra has little overlap, and it is reasonable to set $S_{b} = 0$ for $b \subseteq b_{\rm irr}$. We can then substitute terms into Equation \ref{eq:rt_reformulated} as follows:
\begin{equation}\label{eq:irr_term_weight}
    S_{\rm irr,d,j,b} = -\frac{\kappa_{d,j,b} \rho_{d,j}}{\sum_{l=1}^{n_d}\kappa_{d,l,b} \rho_{d,l}} c\nabla \cdot \vec{F}_{b} \rightarrow S_{\rm irr,d,j} = -c\sum_{b \subseteq b_{\rm irr}} \frac{\kappa_{d,j,b} \rho_{d,j}}{\sum_{j=1}^{n_d}\kappa_{d,j,b} \rho_{d,j}} \nabla \cdot \vec{F}_{b}
\end{equation}
with both the $c$ and $E_{\nu}$ terms multiplying $\kappa_{d,j,b} \rho_{d,j}$ canceling, as they are present in both the numerator and denominator.

Numerically, we evaluate the volume average of  $\nabla \cdot \vec{F}_{b}$ over cell $i$ by making use of the divergence theorem. In 1D Cartesian coordinates, as used in \ref{sec:irradiation_test}, this yields $\left<-\nabla \cdot \vec{F}_{b}\right>_i = -(\Delta x_i)^{-1} (F_b(x_{i+1/2}) - F_b(x_{i - 1/2}))$, where $i - 1/2$ is the left-hand interface and $i + 1/2$ the right-hand interface of the cell. From here, we can discretize the functional form of $F_b$, found in Equation \ref{eq:irr_flux}, to obtain the following:
\begin{equation}
    \left<-\nabla \cdot \vec{F}_{b}\right>_i = \frac{1}{\Delta x_i}\vec{F}_{0, b}\left[\exp\left(-\sum_{i' = i_{\rm beg}}^{i-1}\sum_{j=1}^{n_d} \rho_{d,j} (\kappa_{d, \rm irr})_{j,b} \Delta x_i'\right) - \exp\left(-\sum_{i' = i_{\rm beg}}^{i}\sum_{j=1}^{n_d} \rho_{d,j} (\kappa_{d, \rm irr})_{j,b} \Delta x_i'\right)\right]
\end{equation}
and substituting into Equation \ref{eq:irr_term_weight} yields that
\begin{equation}\label{eq:irr_term_exp_final}
    \left<S_{\rm irr,d,j}\right>_i = \sum_{b \subseteq b_{\rm irr}} \frac{c \kappa_{d,j,b} \rho_{d,j}}{\sum_{j=1}^{n_d}\kappa_{d,j,b} \rho_{d,j}} \left<-\nabla \cdot \vec{F}_{b}\right>_i \,.
\end{equation}

\subsection{Ring-trapping test}\label{sec:ring_trapping}
This test problem is designed to evaluate the implementation of diffusion in curvilinear coordinates, which is particularly relevant to the behavior of dust grains at pressure maxima. The equilibrium between dust-gas drag, which causes drift toward the maximum, and diffusion, which pushes material away, sets the width of the ring in each dust species, and determines whether the species is filtered out by the ring or allowed to pass through. We initialize this problem with gas surface density $\Sigma_g$, dust surface density $\Sigma_d$, and vertically-integrated pressure $p = c_s^2\Sigma$. The sub-Keplerian velocity of this flow is
\begin{equation}
    v_{\phi,g} = \sqrt{v_K^2 + c_s^2 \frac{\partial \ln p_g}{\partial \ln R}} \approx v_K \left(1 + \frac{1}{2} h^2 \frac{\partial \ln p_g}{\partial \ln R}\right) \,.
\end{equation}
The radial drift velocity, $v_{d, r} \approx 2(v_{\phi,g} - v_K)/({\rm St} + {\rm St}^{-1})$, is therefore proportional to the log pressure gradient,
\begin{equation}
\begin{split}
    v_{d,r} &\approx \frac{v_K h^2}{{{\rm St} + {\rm St}^{-1}}}\frac{\partial \ln p_g}{\partial \ln R} = v_{d,r}(R_{\rm max}) + \left.\frac{\partial v_{d,r}}{\partial R}\right|_{R_{\rm max}}(R - R_{\rm max}) + \mathcal{O}((R-R_{\rm max})^2) \,,
\end{split}
\end{equation}
with the second equality arising from Taylor expansion around the point $R = R_{\rm max}$. The first derivative of the drift velocity is
\begin{equation}
    \left.\frac{\partial v_{d,r}}{\partial R}\right|_{R_{\rm max}} = \left(\delta - \frac{1}{2}\right)\frac{v_{d,r}(R_{\rm max})}{R_{\rm max}} + \frac{v_k h^2}{{\rm St} + {\rm St}^{-1}}\frac{\partial}{\partial R}\left(\frac{\partial \ln p_g}{\partial \ln R}\right) \,.
\end{equation}
Setting $R_{\rm max}$ to be a point at which the radial drift velocity---and equivalently, the radial pressure gradient---are zero, we can cancel terms to simplify our expansion to 
\begin{equation}
     v_{d, r} \approx \frac{v_k h^2}{\St + \St^{-1}}\frac{R_{\rm max}}{P_{g, \rm max}}\left(\frac{\partial^2 p_g}{\partial R^2}\right)_{R_{\rm max}}(R - R_{\rm max}) \,.
\end{equation}
In equilibrium, the diffusion velocity $v_{{\rm diff}, d} \equiv -\nu \partial_R \ln(\Sigma_{d}/\Sigma_g)$ is equal and opposite to that of the radial-drift velocity, ensuring that the divergence of the mass flux in the continuity equation is zero. This allows us to solve for the density structure as
\begin{equation}
    \frac{\Sigma_d}{\Sigma_g} = \left(\frac{\Sigma_d}{\Sigma_g}\right)_{R_{\rm max}} \exp \left[\frac{1}{2\alpha ({\St + \St^{-1}})} \left(\frac{1}{p_g}\frac{\partial^2 p_g}{\partial R^2}\right)_{R_{\rm max}} (R - R_{\rm max})^2\right] \,,
\end{equation}
which leads to a characteristic dust concentration width of $w_{dg} \equiv \left[-\alpha ({\St + \St^{-1}})/(p_g^{-1} \partial^2_Rp_g)_{R_{\rm max}}\right]^{1/2} \linebreak = \left[-\alpha ({\St + \St^{-1}})/(R^{-1} \partial_R(\partial \ln p_g/\partial \ln R)_{R_{\rm max}}\right]^{1/2}$, with the latter equality again made under the assumption that $\partial \ln p_g/\partial \ln R = 0$ at $R_{\rm max}$. The implication is that both small, well-coupled ($\St \ll 1$) and large, poorly-coupled ($\St \gg 1$) grains would not concentrate at the pressure bump, being either entrained in the gas or unaffected by it respectively.
\end{appendix}

\end{document}